\documentclass[aps,prb,twocolumn,groupedaddress,showpacs]{revtex4}
\usepackage{graphicx}
\usepackage{dcolumn}
\usepackage{bm}
\usepackage{amsmath,amsfonts,amssymb,amsthm,verbatim}
\usepackage[ansinew]{inputenc}
\usepackage[usenames]{pstcol}

\usepackage{color}
\usepackage{epsfig}

\newcommand{\be}{\begin{equation}}
\newcommand{\ee}{\end{equation}}
\newcommand{\bea}{\begin{eqnarray}}
\newcommand{\eea}{\end{eqnarray}}

\newcommand{\bi}{\begin{itemize}}
\newcommand{\ei}{\end{itemize}}
\newcommand{\bc}{\begin{center}}
\newcommand{\ec}{\end{center}}

\begin{document}

\title{Synthesizing Majorana zero-energy modes in a periodically gated quantum wire}

\author{Mariana Malard$^{1}$, George I. Japaridze$^{2,3}$, and Henrik Johannesson$^{4,5}$}
\affiliation{$\mbox{}^1$University of Brasilia, 70904-910
Brasilia-DF, Brazil}
\affiliation{$\mbox{}^2$Ilia State University, Cholokasvili Ave.
3-5, 0162 Tbilisi, Georgia}
\affiliation{$\mbox{}^3$Andronikashvili
Institute of Physics, Tamarashvili 6, 0177 Tbilisi, Georgia} \affiliation{$\mbox{}^4$Department of
Physics, University of Gothenburg, SE 412 96 Gothenburg, Sweden}
\affiliation{$\mbox{}^5$ Beijing Computational Science Research Center, Beijing 100094, China}

\begin{abstract}

We explore a scheme for engineering a one-dimensional spinless $p$-wave superconductor hosting unpaired Majorana zero-energy modes, using an all-electric setup with a spin-orbit coupled quantum wire in proximity to an $s$-wave superconductor. The required crossing of the Fermi level by a single spin-split energy band is ensured by employing a periodically modulated Rashba interaction, which, assisted by electron-electron interactions and a uniform Dresselhaus interaction, opens a gap at two of the spin-orbit shifted Fermi points. While an implementation in a hybrid superconductor-semiconductor device requires improvements upon present-day capabilities, a variant of our scheme where spin-orbit-coupled cold fermions are effectively proximity-coupled to a BEC reservoir of Feshbach molecules may provide a ready-to-use platform.

\end{abstract}

\pacs{74.78.Fk, 71.10.Pm, 03.67.Lx}

\maketitle

\section{Introduction}

The possible existence of an elementary fermionic particle with the distinguishing property of being its own antiparticle $-$
a Majorana fermion $-$ remains an outstanding puzzle, almost 80 years after the idea was first advanced \cite{Wilczek}. By contrast, {\em emergent}
Majorana fermions are well known to appear in disguise in condensed matter systems $-$ the Bogoliubov quasiparticle in a superconductor being a
notable example  \cite{Chamon,beenakker2014}.

Different, and more intriguing, is the concept of an emergent quasiparticle which is its own antiparticle but exhibits non-Abelian statistics \cite{FreedmanReview} $-$ a {\em Majorana zero-energy mode} (MZM), bound to a defect or a boundary \cite{AliceaReview}. Possible hosts for these particles are fractional quantum Hall systems \cite{MooreRead}, cold gases of fermionic atoms \cite{Sato,jiang2011} and topological superconductors in 1D \cite{Kitaev} and 2D \cite{ReadGreen,Ivanov}. As first realized by Fu and Kane \cite{FuKane1}, the required spinless $p$-wave pairing which makes a superconductor topological may be engineered in a semiconductor structure hybridized with an ordinary $s$-wave superconductor. This has made the topological superconductors the preferred hunting grounds for  MZMs \cite{AliceaReview}, and there is now a variety of theoretical proposals for how to access them in the laboratory. Two schemes, both for proximity-induced 1D $p$-wave pairing, have so far been explored in experiments:  A Rashba spin-orbit coupled quantum wire in proximity to an $s$-wave superconductor and subject to a magnetic field \cite{Sau,Oreg}, and a setup with a chain of magnetic impurities deposited on top of an $s$-wave superconductor \cite{Yazdani}. While the experimental results are promising \cite{SarmaReview}, the verdict is still out as to whether any of them unambigously points to MZMs.

To produce a topological superconducting state in one dimension, the basic trick is to make the Fermi level cross only a single spin-split quasiparticle band. With this, the pairing of the resulting helical (spin-momentum locked) states must then effectively have $p$-wave symmetry so as to make the pair wave function antisymmetric \cite{Kitaev}. In the quantum wire proposals of Refs. \onlinecite{Sau,Oreg}, the trick is carried out by combining a strong Rashba spin-orbit interaction (which causes the spin splitting) with a Zeeman interaction (which pushes one of the bands away from the Fermi level). In the more recent scenario with a magnetic impurity chain on top of an $s$-wave superconductor \cite{Yazdani}, the microscopic spin texture of the chain emulates a combined Rashba and Zeeman interaction to effectively produce a protected set of one-dimensional $p$-wave states in the surface layer of the superconductor. While this latter setup has an advantage in allowing for STM probes of the predicted MZMs, it is more difficult to manipulate and control, and therefore probably less useful for future applications. The quantum wire setup, on the other hand, is easily controllable, with tunable gate voltages that may be used to move around the MZMs in networks of quantum wires $-$ as envisioned in certain architectures for topological quantum gates \cite{Aliceaetal}.

A potential drawback of the quantum wire setup, however, is the reliance on a magnetic field. While the strength of the field can be varied, and allows to tune across the topological quantum phase transition $-$ in this way uncovering experimental signatures of the MZMs $-$ its presence also makes the device less robust against disorder \cite{PotterLee,STS}. Moreover, magnetic fields of the required strengths are difficult to apply locally \cite{SpintronicsReview}, and therefore, integrating them into useful designs for quantum computing with MZMs may prove a challenge. This is particularly so since a universal set of quantum gates \cite{NielsenChuang} using MZMs is obtainable only by supplying ancillary nontopological states \cite{BravyiKitaev,Bravyi}. These states, in turn, may become fragile when subject to a magnetic field. A case in point is when the ancillary states are taken to be spin qubits, as in the proposal in Ref. \onlinecite{LeijnseFlensberg}. To ensure spin degeneracy, the magnetic field must here be precisely tuned, with the spin-up and spin-down states belonging to different orbitals in the quantum dot which hosts them. This sets additional demands on the experimental setup. From a more fundamental point of view, one asks whether there could be a less invasive way to obtain 1D helical electrons (the prerequisite for $p$-wave superconductivity) than breaking time reversal symmetry explicitly, as is the case with magnetic field-based proposals.

In view of this, it is interesting to inquire whether MZMs may be produced in a quantum wire (or network of wires as required for braiding and quantum information processing) using an all-electric scheme, disposing of the magnetic field altogether. In fact, there is already an abundance of theoretical proposals which employ ``non-magnetic" schemes: $d_{x^2-y^2}$ \cite{WongLaw} or $s_{\pm}$-wave proximity pairing \cite{Zhang}, noncentrosymmetric superconductivity \cite{Nakosai}, two-channel quantum wires with channel-dependent spin-orbit interactions \cite{GPF}, or some other mechanism \cite{Chung,Deng,Keselman,Sticlet,Liu,Dumitrescu,Haim,KlinovajaLoss}. Common to these proposals is that they describe quasi-1D (``multichannel") topological superconductors \cite{Schnyder,KitaevClass}, hosting {\em paired} MZMs at each end of the wire (``Majorana Kramers pairs'') \cite{KotetesReview}. In this work we shall instead explore the possibility to generate {\em unpaired} MZMs at the ends of a {\em single-channel} proximity-coupled quantum wire without applying a magnetic field. In our proposed setup, the breaking of time-reversal symmetry (necessary to escape the time-reversal analog of ``fermion doubling" in 1D \cite{konig2008} and obtain helical electron states) is \emph{spontaneous}, and comes about from an interplay between a spatially modulated spin-orbit interaction and the electron-electron (e-e) repulsion. As we shall show, this makes possible a magnetic field-free 1D topological superconductor with a single unpaired MZM at each end of the wire. To the best of our knowledge, schemes for producing unpaired MZMs without the use of a magnetic field has so far been discussed only for especially coupled double-nanowires or multichannel wires \cite{Kotetes}, and for Floquet topological superconductors with a periodic high-frequency driving of the spin-orbit interaction \cite{Reynoso}. Here, we instead make use of a {\em spatially} periodic Rashba spin-orbit interaction in a \emph{single} quantum wire.

Specifically, we shall build on a recent proposal of ours, where a 1D helical system is engineered using a quantum wire subject to a periodically modulated electric field \cite{JJM}. The electric field gives rise to a spatially modulated Rashba spin-orbit interaction, which, when assisted by e-e interactions and a uniform Dresselhaus spin-orbit interaction, opens a gap at two of the spin-orbit shifted Fermi points. As an outcome, a helical Luttinger liquid (HLL) \cite{wu2006,xu2006} emerges at the two remaining gapless Fermi points.  In the present work we inquire about the conditions under which the proximity of an ordinary $s$-wave superconductor could turn this HLL into a 1D spinless $p$-wave superconductor hosting MZMs. The problem becomes nontrivial considering that the induced superconducting pairing competes with the insulating gap-opening process from the modulated Rashba interaction. Using a perturbative renormalization group (RG) argument, we shall find that both processes {\em can} play out concurrently. This establishes a ``proof-of-concept" that a single-channel quantum wire may host unpaired MZMs without the assistance of a magnetic field (or a high-frequency driving of the spin-orbit interaction \cite{Reynoso}). However, a case study with an InAs-based device shows that the required values of the parameters lie outside the experimental range reported for InAs quantum wells. A variant of our scheme with a cold-atom emulation of a quantum wire $-$ where interacting and spin-orbit coupled fermionic atoms are in contact with a BEC reservoir of Feshbach molecules $-$ looks more promising. We shall elaborate on this and argue that a spinless $p$-wave superfluid phase with unpaired MZMs is well supported by a cold atom platform within the parameter regime where our scheme is workable.

The paper is organized as follows. In the next section we introduce a microscopic model for a spin-orbit coupled and periodically gated quantum wire in proximity to an $s$-wave superconductor and advance, through general arguments, that this system may be turned into a spinless $p$-wave superconductor. As mentioned above, the scheme calls for the assistance of e-e interactions and this is discussed in Sec. III through a low-energy effective description of the model. Bosonizing the theory, we then carry out a detailed RG study which allows us to establish the flow equations of the theory in the various parameter regimes. In Section IV we arrive at the phase diagram of the system and provide the minimum practical conditions for sustaining the topological phase in the laboratory. The number of MZMs hosted by the topological superconductor and its possible symmetry classes are discussed in Section V. Finally, in Section VI, we present two case studies - one with a periodically gated InAs quantum wire and the other with an ultracold gas of optically trapped fermionic atoms - intended to assess the experimental viability of our scheme. Our conclusions are given in Section VII.

\section{Synopsis: PHYSICAL PICTURE FROM THE MICROSCOPIC MODEL}

In what follows we present and discuss the microscopic model that captures the physics of the system illustrated in Fig. 1: A quantum wire is gated by a periodic sequence of equally charged top gates and proximity coupled to an $s$-wave superconductor. The electrons in the wire are subject to e-e interactions and two types of spin-orbit interactions: the {\em Dresselhaus}  and {\em Rashba} interactions. The Rashba coupling, being sensitive to an external electric field, will pick up the same modulation of the field from the electrodes. In addition, the chemical potential in the wire gets locally modulated by the electric array. Finally, the superconductor induces $s$-wave pairing in the wire through proximity effect.
\begin{figure}[htpb]
\begin{center}
\includegraphics[scale=0.3]{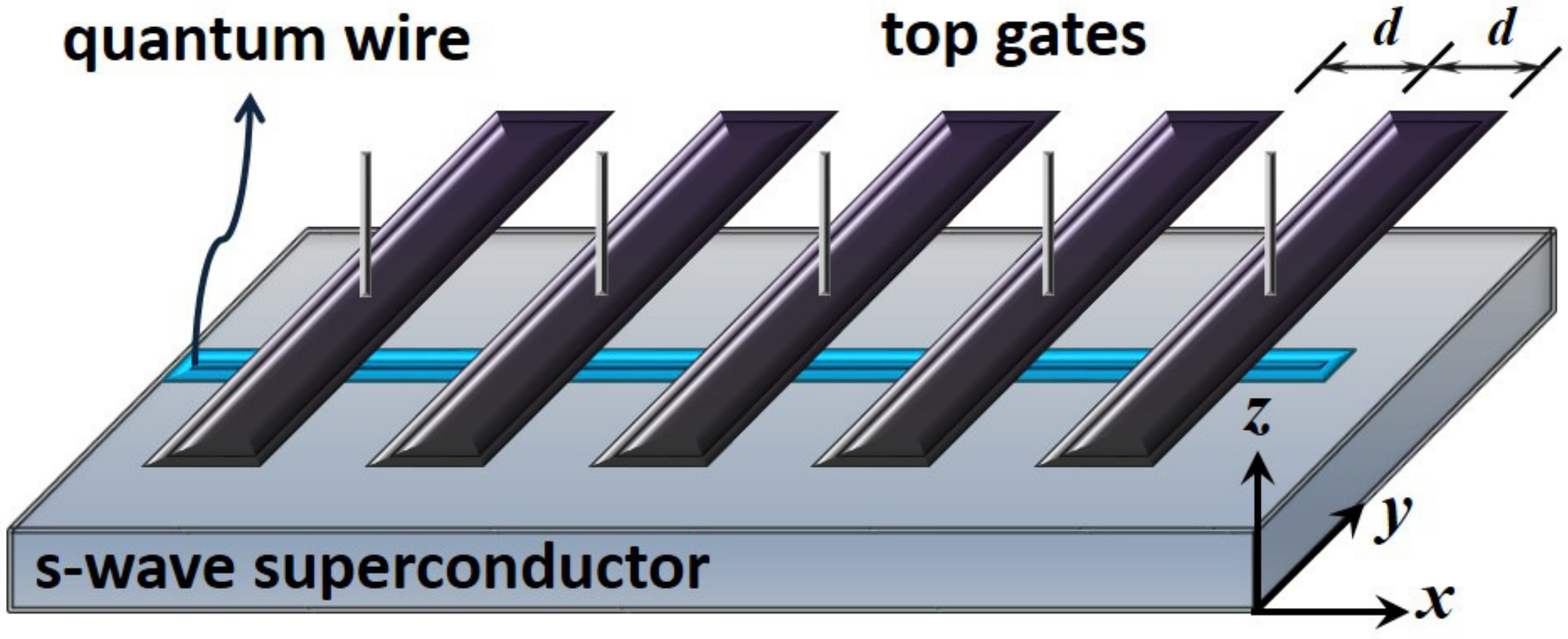}
\caption{(Color online) A quantum wire supporting Rashba, Dresselhaus and e-e interactions is gated by a periodic sequence (with periodicity $2d$) of equally charged top gates. An $s$-wave superconductor induces superconducting pairing in the wire.}
\end{center}
\end{figure}

To better understand the role of the various potentials introduced above, it is instructive to first consider a reduced system obtained by removing the $s$-wave pairing and e-e interactions. The microscopic Hamiltonian thus obtained can be written in a tight-binding formulation as $H=H_{0}+H_{cp}+H_{so}$, with
\begin{eqnarray}
\label{H0} H_{0}\,&=&\,\sum_{n,\alpha}\,[-tc^{\dag}_{n,\alpha}c_{n+1,\alpha}+(\mu/2)c^{\dag}_{n,\alpha}c_{n,\alpha}] \nonumber \\
&-&\!i\!\!\sum_{n,\alpha,\alpha'}\!c^{\dag}_{n,\alpha}[\gamma_{R}\sigma_{\alpha\alpha'}^{y}\!+\!\gamma_{D}\sigma_{\alpha\alpha'}^{x}]c_{n+1,\alpha'} \!+ \!\mbox{H.c.},
\end{eqnarray}
and where
\begin{eqnarray}
\label{HmodC}H_{cp}\!&\!=\!&\!\frac{\mu^{\prime}}{2} \sum_{n,\alpha}
\cos(Qna) c^{\dag}_{n,\alpha}c_{n,\alpha}
\,+\,  \mbox{H.c.}
\end{eqnarray}
\begin{eqnarray}
\label{HmodR} H_{so}\!&\!=\!&\!-i {\gamma'_{R}}\! \sum_{n,\alpha,\beta}\!
\cos(Qna) c^{\dag}_{n,\alpha}\sigma^{y}_{\alpha\beta}c_{n+1,\beta}
+  \mbox{H.c.}
\end{eqnarray}
are the modulated chemical potential and the modulated Rashba spin-orbit interaction, respectively, due to the periodic gating.
Here $c^{\dagger}_{n,\alpha}$ ($c^{\phantom{\dag}}_{n,\alpha}$) creates (annihilates) an electron at site $n$ with spin projection
${\alpha}\!=\,\uparrow,\downarrow$ along the $z$-axis, $t$ is the hopping amplitude, $\mu \, (\mu^{\prime})$ is the amplitude of the uniform (modulated) chemical potential, $\gamma_{D}$ is the amplitude of the uniform Dresselhaus interaction, $\gamma_{R} \, (\gamma'_{R})$ is the amplitude of the uniform (modulated) Rashba interaction (with the former given as a spatial average of the spin-orbit interaction randomized by the ions in nearby doping layers \cite{Sherman,GolubIvchenko}), $\sigma_{\alpha\alpha'}^{x(y)}$ are the matrix elements of the Pauli matrix for the $x \,(y)$-direction, $a$ is the wire lattice spacing and $Q=\pi/d$ is the wave number of the modulation. For a thorough discussion of the modeling described by the Hamiltonian in Eqs. (\ref{H0})-(\ref{HmodR}), we refer the reader to Ref. \onlinecite{MGJJ}.

It is useful to change to a basis that diagonalizes $H_{0}$ in spin space,
\begin{equation}
 \left( \begin{array}{c}
d_{n,+} \\
d_{n,-} \end{array} \right) \equiv \frac{1}{\sqrt{2}} \left(
\begin{array}{c}
-ie^{-i\theta}c_{n,\uparrow}+e^{i\theta}c_{n,\downarrow} \\
e^{-i\theta}c_{n,\uparrow}-ie^{i\theta}c_{n,\downarrow}
\end{array} \right),
\label{newbasis}
\end{equation}
where $\tan(2\theta)=\gamma_{D}/\gamma_{R}$, and $\tau=\pm$ labels the spin projections along the direction of the combined Dresselhaus ($\propto\gamma_{D}\hat{x}$) and uniform Rashba ($\propto\gamma_{R}\hat{y}$) fields. The terms in the Hamiltonian now take the form
\begin{eqnarray}
\nonumber H_{0}&=&\sum_{n,\tau}\,(-t+i\tau\gamma_{eff})d^{\dag}_{n,\tau}d_{n+1,\tau} \\
\label{H0newbasis} &+&\sum_{n,\tau} (\mu/2)d^{\dag}_{n,\tau}d_{n,\tau}\,+ \mbox{H.c.}, \\
\label{HmodCnewbasis}  H_{cp} &=&- \frac{1}{2} \sum_{n,\tau} \mu_n d^{\dag}_{n,\tau}d_{n,\tau} + \mbox{H.c.},  \\
\nonumber H_{so}&=&i\sum_{n,\tau} \cos(2\theta)\,\gamma_{n}\tau d^{\dag}_{n,\tau}d_{n+1,\tau} \\
 \label{HmodRnewbasis} &+&i\sum_{n,\tau}\sin(2\theta)\,\gamma_{n} d^{\dag}_{n,\tau}d_{n+1,-\tau} + \mbox{H.c.},
\end{eqnarray}
where $\gamma_{eff}=\sqrt{\gamma_{R}^{2}+\gamma_{D}^{2}}$, $\mu_{n}=\mu^{\prime}\cos(Qna)$ and $\gamma_{n}=\gamma'_{R}\cos(Qna)$.

The first term, $H_{0}$ in Eq. (\ref{H0newbasis}), can be immediately diagonalized by a Fourier transform, yielding the familiar spin-split spectrum $\varepsilon^{(0)}_{\tau}(k)=-2\tilde{t}\cos(ka-\tau q_{0}a)+\mu$, with $\tilde{t}=\sqrt{t^{2}+\gamma_{eff}^{2}}$ and $q_{0}a=\arctan(\gamma_{eff}/t)$. FIG. 2(a) displays the two lowest $\tau = \pm$ bands inside the first Brillouin zone (BZ). The bands are shifted horizontally by $\pm q_0$, and support four Fermi points $\pm k_F+\tau q_0$ $(\tau = \pm)$, with $k_{F}=\pi N_{e}/2Na$, where $N_e(N)$ is the number of electrons (lattice sites).

Adding the modulated chemical potential term, $H_{cp}$ in Eq. (\ref{HmodCnewbasis}) with the wave number $Q$ written as $Q=(2\pi/a)(p/r)$ for positive integers $p$ and $r$, each band in FIG. 2(a) splits up into $r$ subbands gapped at $\pm k_{r}\equiv\pm m\pi/(ra)=mQ/(2p)$, $m=1,2,...r$, where $\pm k_{r}$ define the boundaries of the new reduced BZs of the periodically gated wire. FIG 2(b) illustrates the six subbands of the case $r=3$ and $p=1$ folded into the first reduced BZ.

By adding also the modulated spin-orbit interaction, $H_{so}$ in Eq. (\ref{HmodRnewbasis}), one may anticipate that its second spin-mixing term (resulting from the interplay between the modulated Rashba and the uniform Dresselhaus interactions as can be seen from the definitions of $\gamma_{n}$ and $\theta$) will lift the degeneracies at the center and at the boundaries of the reduced BZs, through hybridization of  the states with spin projection $\pm$. This is not so, however, since Kramers' theorem forces these states to remain degenerate at the time-reversal invariant points $k=0$ and $\pm k_{r}$. So, while $H_{so}$ will cause some distortion of the subbands in FIG. 2(b), the bands remain connected at $k=0, \pm\pi/3a$.

The picture changes if time-reversal symmetry gets broken, either explicitly (by adding e.g. a magnetic field) or spontaneously, opening a bypass which avoids Kramers' theorem. In this context, recall that strong to intermediate Umklapp scattering in a HLL (which is here suppressed due to the assumed low electron density, being far from half-filling) causes a spontaneous breaking of time-reversal symmetry, with a concurrent opening of a gap in the spectrum \cite{wu2006,xu2006}. As we shall demonstrate, time reversal symmetry similarly gets spontaneously broken when a Coulomb e-e repulsion
\begin{eqnarray}
H_{\text{e-e}}&=&\sum_{n,n',\tau,\tau'}V(n-n')d^{\dag}_{n,\tau}d^{\dag}_{n',\tau'}d_{n',\tau'}d_{n,\tau}
\label{He-enewbasis}
\end{eqnarray}
is added to the Hamiltonian $H=H_{0}+H_{cp}+H_{so}$. In fact, the combined modulated spin-orbit {\em and} e-e- interactions produce a spin-density wave for the electrons at the outer Fermi points, leading to a spontaneous breaking of time-reversal invariance. The presence of the spin-density wave, while being a highly nontrivial phenomenon driven by the collective dynamics, is easy to establish within a bosonization formalism. We will turn to this matter in Sec III.B.

By triggering a spontaneous breaking of time-reversal symmetry in the wire, e-e interactions enable, in effect, the detachment of the bands at the boundaries of a reduced BZ. Specifically, in the next section we show that in the presence of e-e interactions, with the two outer Fermi points residing close to the boundaries of one of the reduced BZs such that $\mid\!Q-2(k_{F}+q_{0})\!\mid \ll {\cal O}(1/a)$ (this will be the first reduced BZ if $p=1$, the second if $p=2$, etc.), gaps open up at these boundaries, lifting the degeneracy of the corresponding states with spin projection $\pm$. As a result, the interior of this reduced BZ will support a HLL at the two remaining gapless Fermi points, with the $\pm$ spin content of these states locked to the direction of motion of the electrons. FIG. 2(c) illustrates the case $p=1$ for which the lowest pair of bands of FIG. 2(b) develop gaps at the zone boundaries after inclusion of e-e interactions.
\begin{figure}[htpb]
\begin{center}

\includegraphics[scale=0.45]{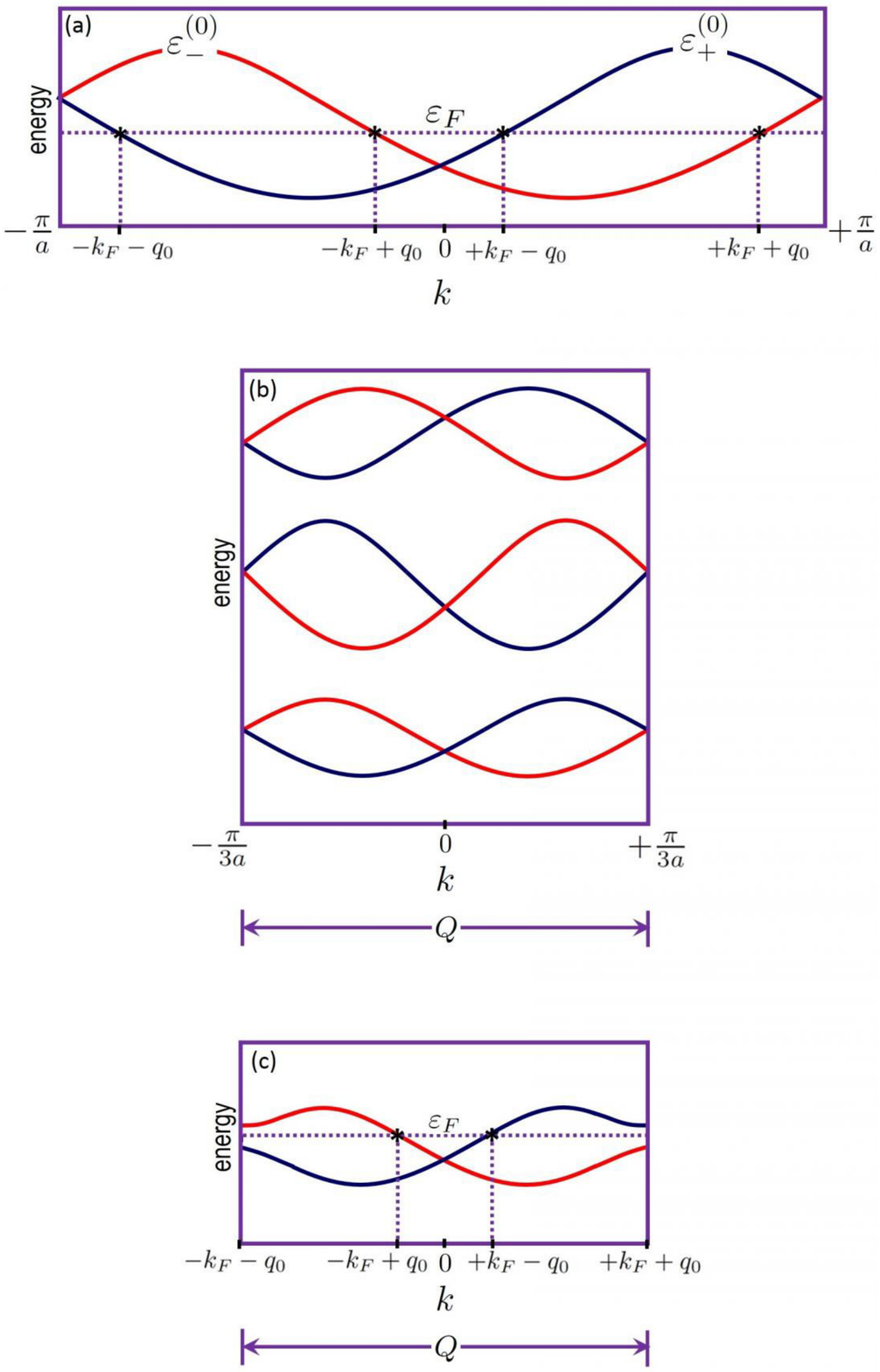}
\caption{(Color online) (a) Lowest bands inside the first BZ of a spin-orbit coupled wire. (b) Subbands inside the first reduced BZ of the periodically gated wire for $r=3$ and $p=1$. (c) Splitting of the two lowest subbands at the zone boundaries in FIG. 2(b), caused by the modulated spin-orbit interaction in the presence of Dresselhaus and e-e interactions (which break time-reversal invariance spontaneously) and provided the outer Fermi points $\pm k_F \pm q_0$ match the boundaries of the first reduced BZ. An HLL emerges in the vicinity of the inner gapless Fermi points $\pm k_F \mp q_0$.}
\end{center}
\end{figure}

To understand how this scenario comes about, and why the interplay between the e-e interaction and the modulated Rashba interaction is key to the process, it is useful to first recall some basics about time-reversal of electron states. For this, let us consider the possibility of a single-particle spin-flip scattering event, from, say, the outer right Fermi point in FIG 2(a), $k_F+q_0$, to the outer left Fermi point, $-k_F -q_0$. Calling the initial and final states $|\alpha\rangle$ and $|\beta\rangle$ respectively, the matrix element for $H$ to connect these states is
\begin{equation} \label{TR1}
\langle \beta | H | \alpha \rangle = \langle \beta | H_{\text{so}} | \alpha \rangle,
\end{equation}
since only $H_{\text{so}}$ in Eq. (\ref{HmodRnewbasis}) can execute a spin flip. With $|\beta\rangle$ being the time-reversed state of $|\alpha\rangle$, $|\beta \rangle =   T |\alpha\rangle$, it follows that
\begin{eqnarray} \label{TR2}
& & \ \ \langle \beta | H_{\text{so}} | \alpha \rangle  \nonumber \\
&=& \ \ \langle T \alpha | H_{\text{so}} | \alpha \rangle = \langle T H_{\text{so}} \alpha | T^2 \alpha \rangle \nonumber \\
&= & -\langle TH_{\text{so}} \alpha | \alpha \rangle
\, = -\langle H_{\text{so}} T \alpha | \alpha \rangle \nonumber \\
&= & -\langle \beta | H_{\text{so}}|\alpha \rangle \nonumber \\
& \implies & \ \ \langle \beta | H_{\text{so}} | \alpha \rangle = 0.
\end{eqnarray}
We have here used that a single-electron state is odd under $T^{2}$, $T^{2} | \alpha\rangle = -| \alpha \rangle$. Also, in the second line of Eq. (\ref{TR2}) we use the anti-unitarity of the time-reversal operator, $\langle T \phi| T \phi' \rangle = \langle \phi'| \phi \rangle$ for any states $|\phi\rangle$ and $|\phi^{\prime}\rangle$, with the identity in the third line following from the time-reversal invariance of $H_{\text{so}}: [H_{\text{so}},T]=0$. Eq. (\ref{TR2}) implies that single-particle spin-flip scattering is forbidden since the matrix element vanishes. However, if   $|\alpha \rangle$ is a two-electron state, one has that $T^2 |\alpha \rangle = |\alpha \rangle$, and it follows that two-particle spin-flip backscattering (from one Fermi point to the opposite) is indeed possible. However, unless the electrons are correlated, the probability that two of them would simultaneously backscatter in response to the spin-flip term in $H_{so}$ is vanishingly small. (In the RG language to be used in Sec. III. C, the process will be demoted to {\em irrelevant}.) This, however, changes when adding the e-e interaction. A two-particle correlated backscattering channel \emph{with spin flip} now opens up, and the process can become {\em relevant} (in the jargon of RG) if the e-e interaction is sufficiently strong. One may, loosely speaking, picture this as resulting from a kind of  ``stimulated" backscattering: Driven by $H_{so}$, an electron may attempt to backscatter with spin flip, but can only do so if it induces, via the e-e interaction, a second electron to do the same. As a result, and as we shall derive formally, the amplitude of the combined process becomes quadratic in the spin-orbit coupling. The modulation of the spin-orbit interaction is crucial for making the process {\em selective}, taking place only for electrons at {\em one} of the pairs of Fermi points, specifically: the outer pair whose separation matches the wave number $Q$ of the external modulation (see FIG. 2(c)). This selectivity is an essential feature of our proposal.

The effective ``spinlessness" of the helical states in the interior of the reduced BZ implies that by incorporating into $H=H_{0}+H_{cp}+H_{so}+H_{\text{e-e}}$ an $s$-wave superconducting pairing potential of strength $\Delta$,
\begin{equation}
H_{sc}\,=\,\sum_{n}\,[\,\Delta\,d_{n,+}d_{n,-}\,+ \mbox{H.c.}\,],
\label{Hscnewbasis}
\end{equation}
may drive the system into a $p$-wave superconducting phase. By means of this, the addition of e-e interactions in effect has triggered a quantum phase transition from an ordinary proximity-coupled $s$-wave superconductor to a topological spinless $p$-wave superconductor, with the ``$p$-waveness"  enforced by the antisymmetry of the pairing wave function that follows from the ``spinless" nature of the helical states. It is interesting to note that this result is anticipated in a work by Stoudenmire {\em et al.} \cite{Stoudenmire}, who hypothesized that a proximity-coupled quantum wire with strong Rashba and Dresselhaus couplings may be driven into a topological phase by interactions, even without an applied magnetic field. In the present work we provide the evidence that this is indeed possible.

The topological nontrivial character of a $p$-wave superconductor \cite{Kitaev} implies that a finite wire, with the charging energy tuned to a degeneracy point \cite{FuTeleportation}, can host localized MZMs $\gamma_{L,i}$ and $\gamma_{R,i}, i=1,2,..,m$ at its left and right ends respectively. These modes, protected by chiral symmetry \cite{TewariSau}, lead to a degenerate groundstate, up to small corrections from wave-function overlaps between left and right MZMs. Specifically, if $\mid\!0\,\rangle$ is a ground state, then $(\gamma_{L,i} + i\gamma_{R,i})\!\!\mid \!0\,\rangle$ is also a ground state, differing from the first by the presence of an extra electron. As we shall see, in the present case there is always only a single unpaired MZM ($m=1$) at each end of the wire.

Having outlined the backbone of our proposal, we close this section by stating the three basic conditions for it to work. First, the proximity gap must be smaller than the dynamically generated insulating gap at the zone boundaries, so that the states of the insulating and empty bands do not mix with the $p$-wave superconducting states. Secondly, the smaller proximity gap must itself exceed the thermal energy so that the device is robust against thermal leakage. Finally, the scaling lengths at which the gaps open up (in the language of RG\cite{Giamarchi}) must fit within the system's cutoff length. The wire has to be sufficiently long also for suppressing the overlap between a left and right MZM wave function (which would otherwise produce a spectral weight for a finite-energy electronic mode). Let us note in passing that having a long wire alleviates the need to build in boundary- and finite-size effects into the description of the HLL. Thus our use of an infinite-volume formalism in what is to come. As we shall see, the conditions above can be given a precise mathematical formulation within the framework of RG. This and other key elements of the theory will be closely examined in what follows.

\section{Low-energy effective theory}

To understand how the modulated spin-orbit interaction, e-e interactions and superconducting paring team up to drive a phase transition from a trivial $s$-wave to a topological $p$-wave superconducting phase in the wire, we shall study the low-energy limit of the full Hamiltonian thus introduced,
\begin{equation}
H=H_{0}+H_{cp}+H_{so}+H_{\text{e-e}}+H_{sc},
\end{equation}
with the terms defined in Eqs. (\ref{H0newbasis})-(\ref{HmodRnewbasis}), (\ref{He-enewbasis}) and (\ref{Hscnewbasis}). We carry out this analysis in three steps: In subsection III.A, we linearize the spectrum around the system's four Fermi points in an extended zone picture, with that producing an effective field theory written in terms of fermionic right- and left-moving field operators. A bosonization procedure is applied in subsection III.B, casting the theory in a form that will be analyzed within an RG formalism in subsection III.C.

\subsection{Linearization of the spectrum}

To fix a working ground (without loss of generality), let us consider the lowest bands of FIG. 2(b). Using a low-energy approach, we will show how these bands can be evolved into the partially gapped band structure depicted in FIG. 2(c). The first step is to choose the Fermi level so that the outer Fermi points $\pm k_{F}\pm q_{0}$ reside in the neighborhoods of the boundaries of the first reduced BZ, so that $\mid\!Q-2(k_{F}+q_{0})\!\mid \ll {\cal O}(1/a)$. (For ease of exposition we put $Q=2(k_{F}+q_{0})$ in the following. However, all results obtained in the continuum limit remain valid as long as $\mid\!Q-2(k_{F}+q_{0})\!\mid \ll {\cal O}(1/a)$.) Here it is important to note that the modulation wave number $Q$ is likely to be preset in an experimental device. Thus, rather than choosing $Q$, it is instead $k_{F}$ that is tuned $-$ by filling up the system via a backgate $-$ so as to make the outer Fermi points approach the zone boundaries.

Having thus defined the Fermi level, the next step is to linearize the spectrum around the four Fermi points $\pm k_{F}+\tau q_{0}$ ($\tau=\pm$). This calls for an extended zone scheme that takes advantage of translational symmetry to formally ``disentangle" the bands at the boundaries of the reduced BZ. This scheme is represented in FIG. 3(a)-(b): the enclosed pieces of the $+$ and $-$ bands are displaced by the reciprocal vector $Q$, rendering two parabolic-like bands in the extended zone scheme. With this, the linearization of the spectrum, as illustrated in FIG. 3(c), can be carried out in the standard way.
\begin{figure}[htpb]
\begin{center}
\includegraphics[scale=0.4]{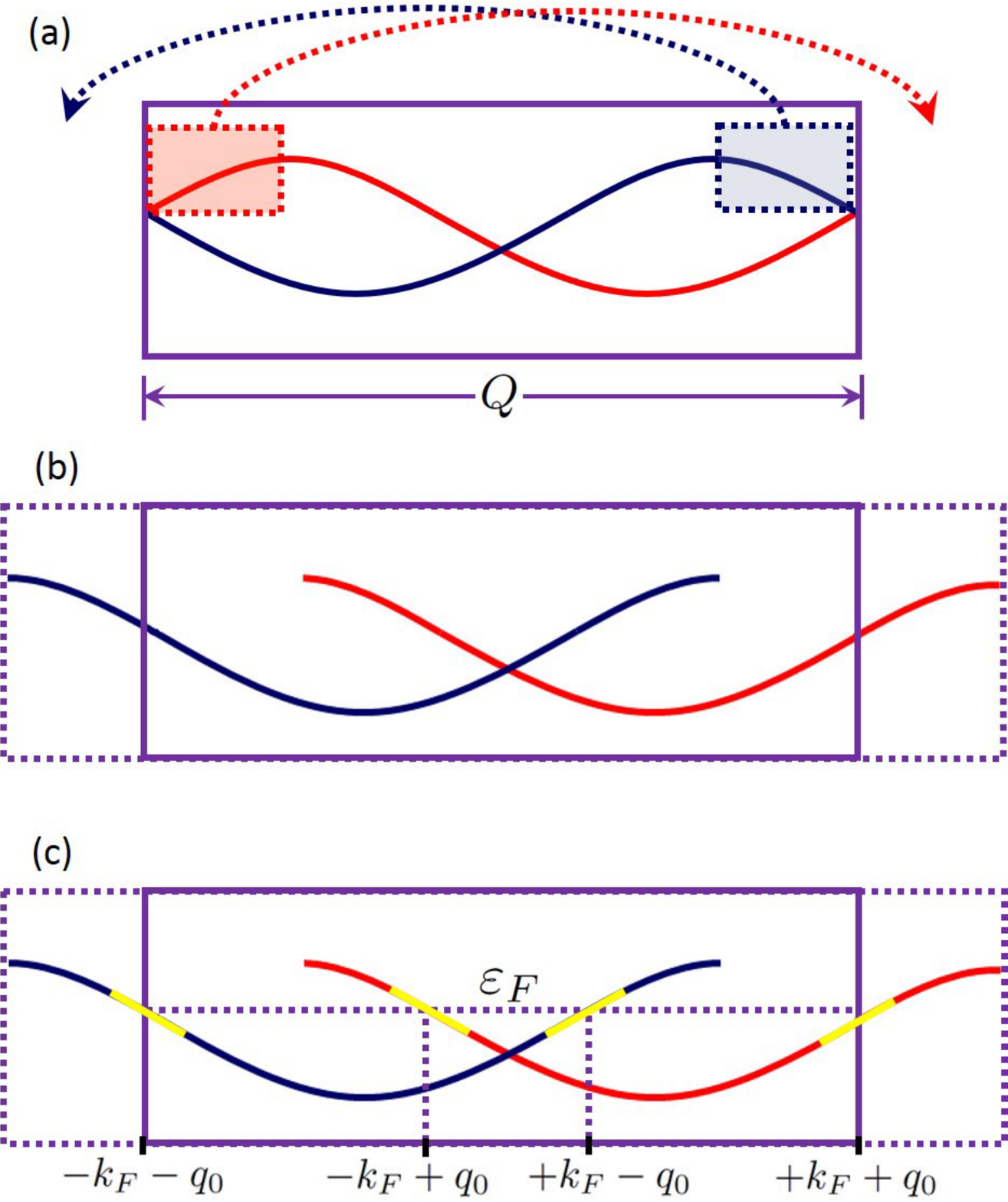}
\caption{(a) Energy bands (blue for $+$ and red for $-$) with boxed segments being displaced by a reciprocal vector. (b) Resulting spin-split parabolic-like bands in the extended zone scheme. (c) Linearization (yellow segments) of the spectrum around the Fermi level.}
\end{center}
\end{figure}

The continuum limit of the low-energy (linearized) theory is obtained through the transformations $na\rightarrow x$, $\sum_{n}\rightarrow\int dx/a$ and
\begin{equation}
\nonumber  d_{n,\tau}\rightarrow\sqrt{a}(e^{i(+k_{F}+\tau q_{0})x}
R_{\tau}(x)+e^{i(-k_{F}+\tau q_{0})x}L_{\tau}(x)),
\end{equation}
where $R_{\tau}(x)$ and $L_{\tau}(x)$ are fermionic field operators that annihilate right- and left-moving excitations at the respective Fermi points. Specifically, $L_{-}$ and $R_{+}$ apply to the ``outer" Fermi points $-k_{F}-q_{0}$ and $+k_{F}+q_{0}$ respectively, while $L_{+}$ and $R_{-}$ apply to the ``inner" ones, $-k_{F}+q_{0}$ and $+k_{F}-q_{0}$ respectively.

Omitting rapidly oscillating terms that vanish upon integration when $Q=2(k_{F}+q_{0})$, we find that $H \rightarrow \int dx\,({\cal H}_{outer}+{\cal H}_{inner}+{\cal H}_{e-e})$ where
\begin{eqnarray}
\label{Houter}
\nonumber {\cal H}_{outer}&=&-iv_F(:\!R^{\dag}_{+}\partial_{x}R_{+}\!:\,-\,:\!L^{\dag}_{-}\partial_{x}L_{-}\!:)\, \\
&+&\lambda(R^{\dag}_{+}\partial_{x}L_{-}\,+\,L^{\dag}_{-}\partial_{x}R_{+}) \\
\nonumber &+&\,\Delta(L^{\dag}_{-}R^{\dag}_{+}\,+\,R_{+}L_{-}), \\
\label{Hinner}
\nonumber{\cal H}_{inner}\,&=&\!-iv_F(:\!R^{\dag}_{-}\partial_{x}R_{-}\!:\,-\,:\!L^{\dag}_{+}\partial_{x}L_{+}\!:) \\
&+&\,\Delta(R^{\dag}_{-}L^{\dag}_{+}\,+\,L_{+}R_{-}), \\
\nonumber{\cal H}_{\text{e-e}}&=&\sum_{\tau,\tau'}\,g_{1}\!:\!R^{\dag}_{\tau}L_{\tau}L^{\dag}_{\tau'}R_{\tau'}\!:
+ \, {g}_{2}\!:\!R^{\dag}_{\tau}R_{\tau}L^{\dag}_{\tau'}L_{\tau'}\!: \\
&+&\,\frac{g_{2}}{2}(:\!L^{\dag}_{\tau}L_{\tau}L^{\dag}_{\tau'}L_{\tau'}\!:\,+\,L\rightarrow R),
\label{densityHe-e}
\end{eqnarray}
with $v_{F}=2a\tilde{t}\sin(k_{F}a)$, $\lambda=a\gamma'_{R}\gamma_{D}/\gamma_{eff}$, $g_{1}\sim\tilde{V}(k\sim2k_{F})$,
$g_{2}\sim\tilde{V}(k\sim0)$, $\tilde{V}(k)$ being the Fourier transform of the Coulomb potential, and where $:...:$ denotes normal ordering. The backscattering process $\sim g_1$ is known to be RG irrelevant in a Luttinger liquid \cite{Giamarchi}, and the same holds true in the presence of spin-orbit interactions and superconducting pairing \cite{Schulz}. Here we thus consider only the dispersive and forward scattering processes $\sim g_2$ in Eq. (\ref{densityHe-e}). The irrelevant $g_{1}$-backscattering, which conserves spin, is not to be confused with the spin-flip backscattering discussed in Sec. II. The latter is a two-particle correlated backscattering induced by the modulated spin-orbit interaction which, as we shall show, becomes relevant in the presence of strong enough $g_{2}$-processes.

\subsection{Bosonized theory}

The physics of Eqs. (\ref{Houter})-(\ref{densityHe-e}) reveals itself in a more transparent way if we go to a bosonized picture by applying the prescription \cite{Giamarchi}:
\begin{equation}
R_{\tau}\,=\,\frac{\eta^{R}_{\tau}}{\sqrt{2\pi
a}}e^{-i\sqrt{\pi}(\phi_{i}+\tau\theta_{i})},
\label{R}
\end{equation}
\begin{equation}
L_{\tau}\,=\,\frac{\eta^{L}_{\tau}}{\sqrt{2\pi
a}}e^{+i\sqrt{\pi}(\phi_{j}+\tau\theta_{j})}.
\label{L}
\end{equation}
Here $i=1\,(2)$ applies to $\tau=+\,(-)$ and $j=1\,(2)$ applies to $\tau=-\,(+)$; $\phi_{i}(x)$ and $\theta_{i}(x)$ are dual bosonic fields satisfying $v_{F}\partial_{x}\theta_{i}=\mp\partial_{t}\phi_{i}$ (with $i=1\,(2)$ for the minus (plus) sign), and $\eta_{\tau}^{R,L}$ are the Klein factors needed to preserve the Fermi statistics of the $R_{\tau}$ and $L_{\tau}$ fields.

The bosonized Hamiltonian reads $H=\int dx\, ({\cal H'}_{outer}+{\cal H'}_{inner}+{\cal H}_{mix})$ with
\begin{eqnarray}
\label{Houterboson} \nonumber {\cal H'}_{outer}&=&u[(\partial_{x}\theta_{1})^{2} \!+\! (\partial_{x}\phi_{1})^{2}]
- \frac{\Delta}{\pi a}\sin(\sqrt{\frac{4\pi}{K}}\theta_{1}) \\
 &+&\frac{\lambda}{\sqrt{\pi K}a}\cos(\sqrt{4\pi K}\phi_{1})\partial_{x}\theta_{1}, \\
 \label{Hinnerboson} {\cal H'}_{inner}&=&\! u[(\partial_{x}\theta_{2})^{2}\!+\!(\partial_{x}\phi_{2})^{2}]\!-\!\frac{\Delta}{\pi a}\sin(\sqrt{\frac{4\pi}{K}}\theta_{2}), \\
\label{Hmix} {\cal H}_{mix}&=&\frac{g_{2}K}{\pi}\partial_{x}\phi_{1}\partial_{x}\phi_{2},
\end{eqnarray}
where $K=(1+g_{2}/(\pi v_{F}))^{-1/2}$ is the Luttinger parameter, and $u=v_{F}/2K$ is the Fermi velocity dressed by e-e interactions. The non-interacting limit corresponds to $K=1$ (i.e. $g_2=0$), for which ${\cal H}_{mix} =0$, and, referring back to Eqs. (\ref{Houter}) and (\ref{Hinner}), ${\cal H}^{\prime}_{outer} = {\cal H}_{outer}$ and ${\cal H}^{\prime}_{inner} = {\cal H}_{inner}$. The bosonized theory is thus seen to split into two branches given by ${\cal H'}_{outer}$ and ${\cal H'}_{inner}$, each acting at the corresponding pair of outer and inner Fermi points, and, for $K \neq 1$, coupled by the density-density interaction ${\cal H}_{mix}$.

In Ref. \onlinecite{JJM} we analyzed the bosonized theory defined by Eqs. (\ref{Houterboson})-(\ref{Hmix}) in the absence of superconducting pairing, i.e. with $\Delta=0$. Going to a path integral formulation, we found that by integrating out the last term in Eq. (\ref{Houterboson}), the outer branch gets described by a quantum sine-Gordon model with potential $\propto\lambda^{2}\cos(\sqrt{16\pi K}\phi_{1})$. This potential is strongly RG-relevant if $K<1/2$ (strong e-e repulsion). If $K\geq1/2$ (weak e-e repulsion), it is marginally RG-relevant provided the strength $\lambda$ of the modulated spin-orbit interaction is sufficiently large, satisfying $(\lambda/v_{F})^{2}>(2-1/K)$ \cite{mariana2013}. In both cases, the term opens a gap for the electrons in the outer branch, at the same time as it suppresses the branch-mixing term in Eq. (\ref{Hmix}) by pinning the $\phi_{1}$-field. As a result, the inner branch decouples and comes to support an HLL. Folding back the extended zone into the first reduced BZ, we arrive at the gapped band structure anticipated in FIG. 2(c). If $K\geq1/2$ but $(\lambda/v_{F})^{2}\leq(2-1/K)$, the spin-orbit potential becomes RG-irrelevant, the bands remain gapless, and the system an ordinary Luttinger liquid.

The HLL put forward in Ref. \onlinecite{JJM} is different from the ones that have so far been studied experimentally: It is neither holographic \cite{konig2008} (unlike the edge states of a quantum spin Hall insulator) nor quasihelical \cite{braunecker2013} (unlike a magnetic-field-assisted helical liquid). The time-reversal analog of the fermion-doubling problem implied by Kramers' theorem \cite{konig2008} is instead avoided by the fact that the gapped branch breaks time-reversal symmetry {\em spontaneously} by developing a spin-density wave (SDW). This can be seen from an analysis in Ref. \onlinecite{wu2006}, which, when carried over to ${\cal H'}_{outer}$ in Eq. (\ref{Houterboson}) with $\Delta=0$, reveals that the Ising-like SDW operator $i(R^{\dagger}_{+}L_{-}-\mbox{H.c.})\sim\cos(\sqrt{4\pi}\phi_1)$ takes on a finite expectation value in the gapped ground state (due to the pinning of the $\phi_1$-field) \cite{JJM}. Is is important to point out that this spontaneous breaking of time reversal symmetry (that enables the modulated spin-orbit interaction to gap out one branch, isolating a HLL in the other) is only possible in the presence of sufficiently strong e-e interactions. If the interaction is weak ($K\approx1$), a pinning of $\phi_{1}$ would require the marginal RG-flow to be launched from an impracticably high point in the $K-\lambda$ plane with $\lambda>\sqrt{2}v_{F}$.

Disregarding, for the moment, the coupling between the two branches given by Eq. (\ref{Hmix}), let us now consider the effect of the superconducting pairing. Switching on the pairing field $\Delta$, the inner branch acquires a sine-Gordon term as given by Eq. (\ref{Hinnerboson}). This perturbation is strongly RG-relevant if $K>1/2$ (weak e-e repulsion), while if $K\le1/2$ (strong e-e repulsion) it is marginally relevant provided the strength $\Delta$ of the superconducting pairing is enough to survive the e-e repulsion, satisfying $a\Delta/v_{F}>(1/K-2)$. In both cases, a superconducting gap opens up in the inner branch. If, on the other hand, $K\le1/2$ with $a\Delta/v_{F}\le(1/K-2)$, the superconducting pairing gets suppressed by strong e-e repulsion, becoming RG-irrelevant.

One can now envision that combining a marginally/strongly relevant superconducting pairing in the inner branch with a strongly/marginally relevant spin-orbit interaction in the outer branch, the $s$-wave-coupled helical electrons will undergo a transition to a $p$-wave topological phase. Because the outer branch is now also subject to superconducting pairing (see Eq. (\ref{Houterboson})), the parameter regime within which this phase transition takes place depends on how superconductivity and spin-orbit coupling play out together in that branch. Moreover, reinstating the branch-mixing term, Eq. (\ref{Hmix}), the emergence of a topological phase is conditioned to $H_{mix}$ becoming dynamically frozen out on the relevant length scale so that the inner and outer branches become effectively decoupled. As we shall argue, inside a properly chosen parameter regime, the opening of an insulating gap in the outer branch precisely provides for this. The task in hand is, therefore, to determine this parameter regime in which the insulating order dominates superconducting pairing correlations in the outer branch, and hence makes possible the emergence, in the inner branch, of a decoupled helical state that will go topological under superconducting pairing.

\subsection{Renormalization Group analysis of the outer branch}

We start by examining the competition between the spin-orbit interaction and the superconducting pairing in the outer branch by investigating the theory given by Eq. (\ref{Houterboson}) as such, not concerning ourselves, in this Section, with the effects from the mixing between the two branches in Eq. (\ref{Hmix}).

For this purpose, it is convenient to go to a Lagrangian formalism. After carrying out a Legendre transformation of Eq. (\ref{Houterboson}), and integrating out the conjugated momentum field $\Pi_{1}=-\partial_{x}\theta_{1}$ from the partition function, we arrive at the effective action for the outer branch:
\begin{multline} \label{Action}
S_{outer}= \int dxd\tau \Big[ \frac{2u}{2}\Big((\partial_{x}\phi_{1})^{2}+\frac{1}{(2u)^{2}}(\partial_{\tau}\phi_{1})^{2}\Big) \\
- \! \frac{v_{F}g_{so}}{\pi a^{2}}\cos(\sqrt{16\pi K}\phi_{1})-\frac{v_{F}g_{sc}}{\pi a^{2}}\cos(\sqrt{\frac{4\pi}{K}}\theta_{1}) \Big],
\end{multline}
where $\tau=it$ is the imaginary time and where $g_{sc}=a\Delta/v_{F}$ and $g_{so}=\lambda^{2}/(4v_{F}^{2})$ are dimensionless coupling constants. Note that the spin-orbit coupling $g_{so}$ is quadratic in the amplitude of the modulated Rashba interaction, as anticipated from our qualitative discussion of the correlated backscattering in Sec. II.

The action (\ref{Action}) is an extended version of the sine-Gordon model where, besides the usual mass term given by the cosine of the $\phi_{1}$-field, a cosine of the dual $\theta_{1}$-field is also present. This model has been a subject of intensive studies during the past decades \cite{Jose,Boyanovsky,Lecheminant}. We note the manifest invariance of Eq. (\ref{Action}) under the duality transformation $\phi_{1}\leftrightarrow\theta_{1}$ and $2K\leftrightarrow1/(2K)$ when $g_{so}=g_{sc}$, i.e. the property of a self-dual sine-Gordon model. For details we refer the reader to the Ref. [\onlinecite{Gogolin}].

A crucial feature of the model described by Eq. (\ref{Action}) is that its two cosine potentials are mutually nonlocal and, therefore, cannot be minimized simultaneously \cite{Lecheminant}. This property per se suggests that the theory must support two regimes, each governed by one of the antagonistic spin-orbit and superconducting terms. But the outcome of the competition between the two regimes depends not only on the relation between the corresponding energy scales $g_{so}$ and $g_{sc}$, but also on the energy scale of the e-e interaction as given by the Luttinger parameter $K$. In fact, the scaling dimensions $\Delta_{so}$ and $\Delta_{sc}$ of the spin-orbit and superconducting perturbation, respectively, are controlled by the e-e interaction: $\Delta_{so}=4K$ and $\Delta_{sc}=1/K$. A necessary condition for a perturbation to be strongly relevant is that its scaling dimensionality be less than 2, else the perturbation will be irrelevant or, at most, marginally relevant. Therefore, since $\Delta_{so}\Delta_{sc}=4$, when either one of the cosine perturbations is strongly relevant, then the other perturbation must be irrelevant or marginally relevant.

When one of the perturbations is strongly or marginally relevant and the other is irrelevant, the low-energy physics of the model is simply governed by the relevant operator and the problem effectively reduces to the standard sine-Gordon model, either for the $\phi_{1}$-field or for the $\theta_{1}$-field. In this case, the resulting low-energy theory is fully massive and the continuous translational symmetry of the free gapless Gaussian model is broken down to the discrete $Z_N$ symmetry associated with the minima of the relevant cosine term. The corresponding field becomes pinned in one of these minima. In the case of the $g_{so}$-perturbation, the symmetry breaking is translated into an insulating order sustained by a dynamically generated soliton gap \cite{Gogolin}, whereas for the $g_{sc}$-perturbation, the symmetry breaking is translated into a superconducting order sustained by a superconducting gap.

To uncover this process in detail, we exploit the perturbative RG solution of the model given by Eq. (\ref{Action}) obtained using an operator product expansion of the S-matrix. Defining the electron-electron interaction parameter $g_{ee}$ through $K\approx1/2-g_{ee}$, the RG flow equations for $g_{ee}$, $g_{so}$ and $g_{sc}$ can be read off from Ref. \onlinecite{Gogolin}:
\begin{equation}
\frac{dg_{ee}}{dl}=g_{so}^{2}-g_{sc}^{2}\,,
\label{gppflow}
\end{equation}
\begin{equation}
\frac{dg_{so}}{dl}=4g_{so}g_{ee}\,,
\label{gsoflow}
\end{equation}
\begin{equation}
\frac{dg_{sc}}{dl}=-4g_{sc}g_{ee}\,,
\label{gscflow}
\end{equation}
where $l=\ln s$, with $s$ a scale factor.

By numerically solving these equations we obtain the RG flows $g_{ee}(l)$, $g_{so}(l)$ and $g_{sc}(l)$ of the corresponding parameters in the outer branch. FIG. 4 displays the resulting phase diagram for different sets of $g_{ee}$, $g_{so}$ and $g_{sc}$ bare ($l=0$) values. For better visualization, we {have split} the phase diagram in two separate panels - FIGS. 4(a) and 4(b) - according to the sign of the bare $g_{ee}$.
\begin{figure}
\includegraphics[scale=0.5]{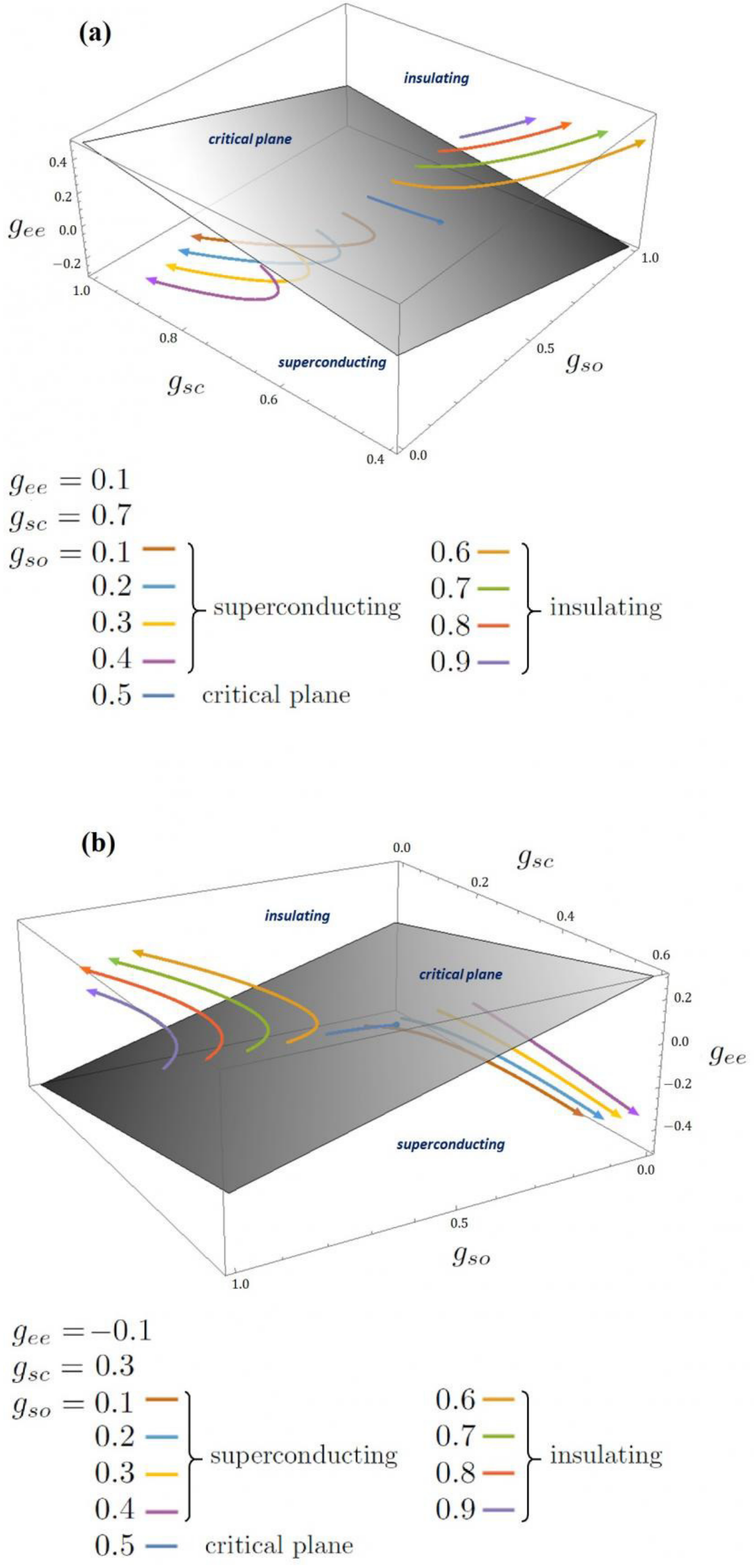}
\caption{(Color online) RG phase diagram of the outer branch obtained by numerical solution of the flow equations (\ref{gppflow})-(\ref{gscflow}). The critical plane equation is: $g_{so}-g_{sc}+2g_{ee}=0$. (a) bare $g_{ee}>0$ (b) bare $g_{ee}<0$}
\end{figure}

FIG. 4 shows that the phase diagram of the outer branch consists of two regions separated by a critical plane which is the locus of the theory's fixed points. The plane equation obtained from the numerics
\begin{equation}
g_{so}-g_{sc}+2g_{ee}=0
\label{criticalplane}
\end{equation}
can also be derived analytically, as shown in Ref. \onlinecite{Gogolin}. For initial values of the parameters corresponding to a point on the plane, the resulting flow will be constrained to the plane, eventually sticking to a fixed point.

Below the critical plane, the spin-orbit interaction becomes irrelevant whereas the superconducting pairing becomes \emph{marginally} relevant if the bare $g_{ee}\geq0$ (FIG. 4(a)) and \emph{strongly} relevant if the bare $g_{ee}<0$ (FIG. 4(b)). As a result, a pairing gap opens below the plane, leading to a superconducting phase in the outer branch. More interesting, for the realization of our scheme, is the region above the critical plane. Here superconducting pairing goes irrelevant, while the spin-orbit potential becomes \emph{strongly} relevant if the bare $g_{ee}>0$ (FIG. 4(a)) and \emph{marginally} relevant if the bare $g_{ee}\leq0$ (FIG. 4(b)). It follows that an insulating gap opens up above the plane, sustaining an insulating phase in the outer branch.

\section{Phase diagram}

To delineate the phase diagram of the system, we now combine the parameter regimes discussed in Sec III.B for the inner branch (rewritten in terms of the parameters $g_{ee}$ and $g_{sc}$) with the regimes obtained from the analysis carried out in Sec. III.C for the outer branch. Here, the coupling between the branches must be addressed in order to correctly characterize the emerging phases.

Recall from the analysis in Sec. III.C, carried out in the absence of the branch-mixing term Eq.(\ref{Hmix}), that when the spin-orbit interaction becomes strongly or marginally relevant in the outer branch the associated $\phi_{1}$-field gets pinned. Since the branch-mixing term is marginal (has scaling dimension equal to 2), in the presence of a strongly relevant spin-orbit interaction it gets suppressed by the pinning of $\phi_{1}$ already at a short length scale (short in the RG sense, that is: shorter than the scale at which the branch-mixing would start to affect the RG flow of the spin-orbit and pairing interactions in a consequential way). As a result, the inner and outer branches decouple above the critical plane in FIG. 4(a). On the other hand, above the critical plane in FIG. 4(b), it is possible that the flow of the only marginally relevant spin-orbit interaction will get distorted by the (also marginal) branch-mixing term in such a way as to ultimately prevent the pinning of $\phi_{1}$, in which case the branch-mixing would survive and the branches would remain coupled. More opportune for our purpose would be if the distortion on the marginally relevant spin-orbit flow would not halt the pinning of $\phi_{1}$, thus preserving the branch decoupling. We shall return to this point below. Finally, if the superconducting pairing is the strongly or marginally relevant operator, then the pinned field would be $\theta_{1}$, with no suppression effect upon $H_{mix}$. Therefore, in this case, the branches remain coupled all across the region below the critical plane in FIGS. 4(a) and 4(b).

The table in FIG. 5 combines the parameter regimes of the outer and inner branches and shows the resulting phases, characterized as ``branch-coupled", ``branch-decoupled" or ``unknown" according to the discussion above.
\begin{figure}
\includegraphics[scale=0.4]{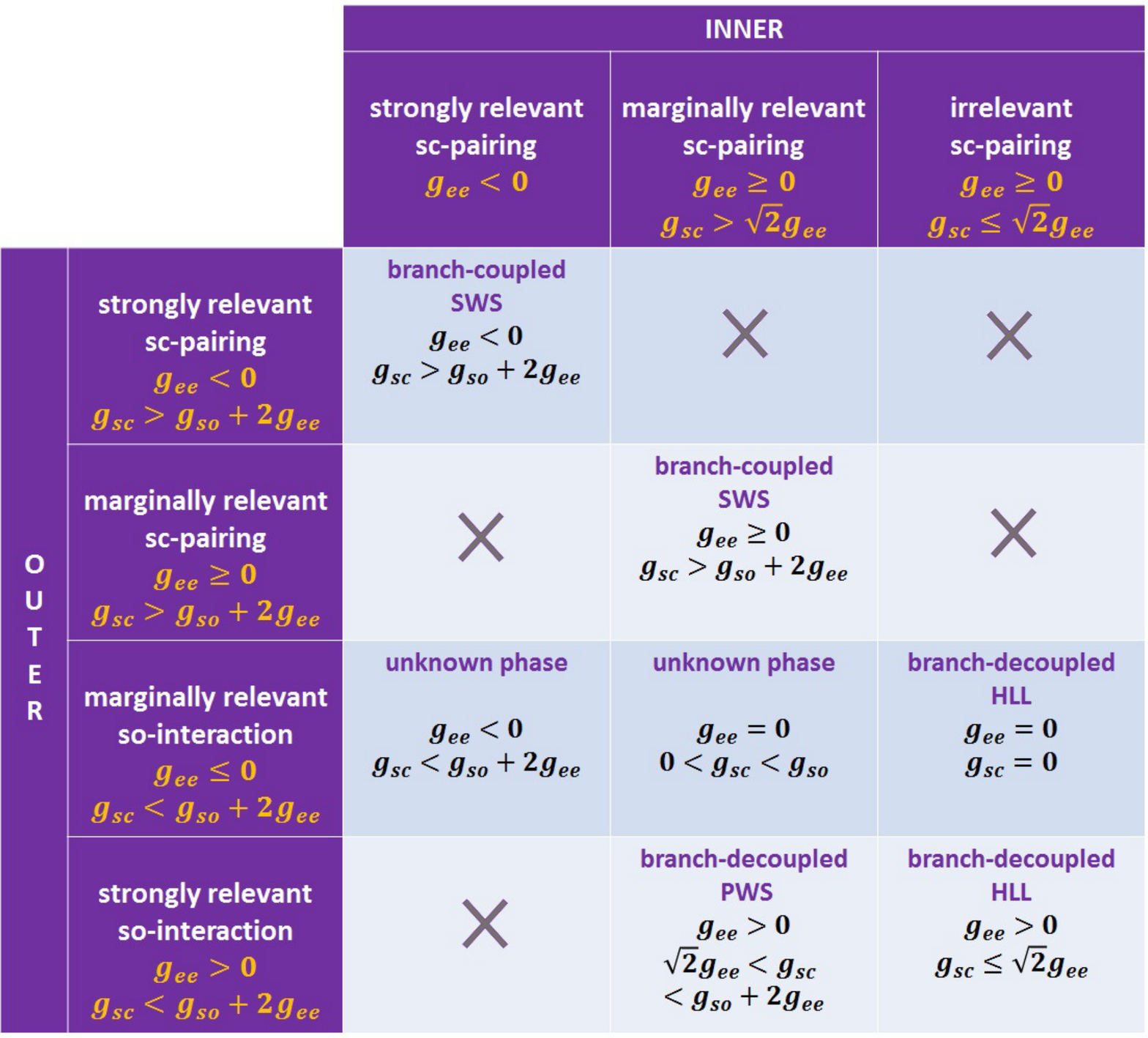}
\caption{(Color online) Combining the parameter regimes of the outer and inner branches, taking into account the role of the coupling between the branches in the characterization of the emerging phases. The abbreviation SWS is short for $s$-wave superconductor, PWS is for $p$-wave superconductor and, as before, HLL is for helical Luttinger liquid.}
\end{figure}

The ``branch-decoupled PWS" is our target phase: the $p$-wave paired topological superconductor hosting MZMs at its ends. In the ``branch-decoupled HLL" entries, the system simply reduces to the HLL realization when the superconducting pairing becomes irrelevant in the inner branch (entry 4-3) or is absent (entry 3-3). (At first sight, the ``branch-decoupled" characterization in entry 3-3 may appear contradictory with the discussion above since the corresponding state - a single point in the phase diagram - arises from a marginally relevant spin-orbit interaction. However, the competing superconduting pairing being simply absent from this state ($g_{sc}=0$ in this ``pure" HLL realization), the spin-orbit interaction, even if only marginally relevant, will eventually pin the $\phi_{1}$-field in the outer branch, suppressing the branch-mixing.)

The two ``branch-coupled SWS" phases in FIG. 5 are simple to assess. In these cases, the spin-orbit interaction has been washed out from the outer branch (is RG irrelevant) and the full theory reduces to two identical sine-Gordon models, one for each branch, coupled by the branch-mixing term: Eqs. (\ref{Houterboson})-(\ref{Hmix}), with $\lambda\rightarrow0$. Since there is no distinction between the outer and inner branches in these phases, the branch-mixing term Eq. (\ref{Hmix}) can actually be absorbed by a simple rotation to a basis in which the system is not resolved in terms of inner and outer branches, but is described in terms of collective excitations with no definite helicity. For example, one may rotate the $\{\phi_{1},\phi_{2}\}$-(inner and outer basis) to the standard $\{\phi_{\rho},\phi_{\sigma}\}$-(charge and spin basis). The result is an $s$-wave superconductor with pairing amplitude dressed by the e-e forward scattering. The $s$-wave superconductors of entries 1-1 and 2-2 differ from each other in that, in the first case (strongly relevant superconducting pairing), the pairing gap opens up and stabilizes the $s$-wave phase at a shorter length scale than in the second case (marginally relevant superconducting pairing).

Finally, in the ``unknown" phases of FIG. 5 the underlying physics does not surface from a simple rotation of basis. In these phases, the selectiveness of the modulated spin-orbit interaction (which acts, as we have seen, only on the external Fermi points whose separation matches the wave number $Q$ of the external modulation) leads to the differentiation of the outer and inner branches. This separation demands a choice of basis - our outer and inner basis - capable of resolving the helical nature of the system. The cost of this basis is the presence of the branch coupling $H_{mix}$ in Eq. (\ref{Hmix}). This coupling simply encodes the Coulomb forward scattering process connecting electrons from the outer and inner branches, with equal chirality. The treatment of $H_{mix}$, out of the regime where this term is suppressed, calls for methods that go beyond the present bosonization-perturbative RG approach. To explore whether the combination of a marginally relevant spin-orbit interaction in the outer branch and a strongly or marginally relevant superconducting pairing in the inner branch may in fact be sufficient to generate the desired topological $p$-wave phase is an interesting problem which we expect could be analyzed using DMRG or some other numerical approach.

The phase diagram of the system that transpires from our analysis is depicted in FIG. 6 (a fixed-$g_{so}$ cut through the 3D phase diagram). We should stress that this phase diagram is a first order approximate prediction, obtained by extrapolating the RG equations (\ref{gppflow})-(\ref{gscflow}) to the entire $[-1/2,+1/2]$ range of $g_{ee}$. However, away from $g_{ee}\approx0$ higher-order corrections may enter into the RG equations, affecting the length scale of gap opening within each phase.
\begin{figure}
\includegraphics[scale=0.32]{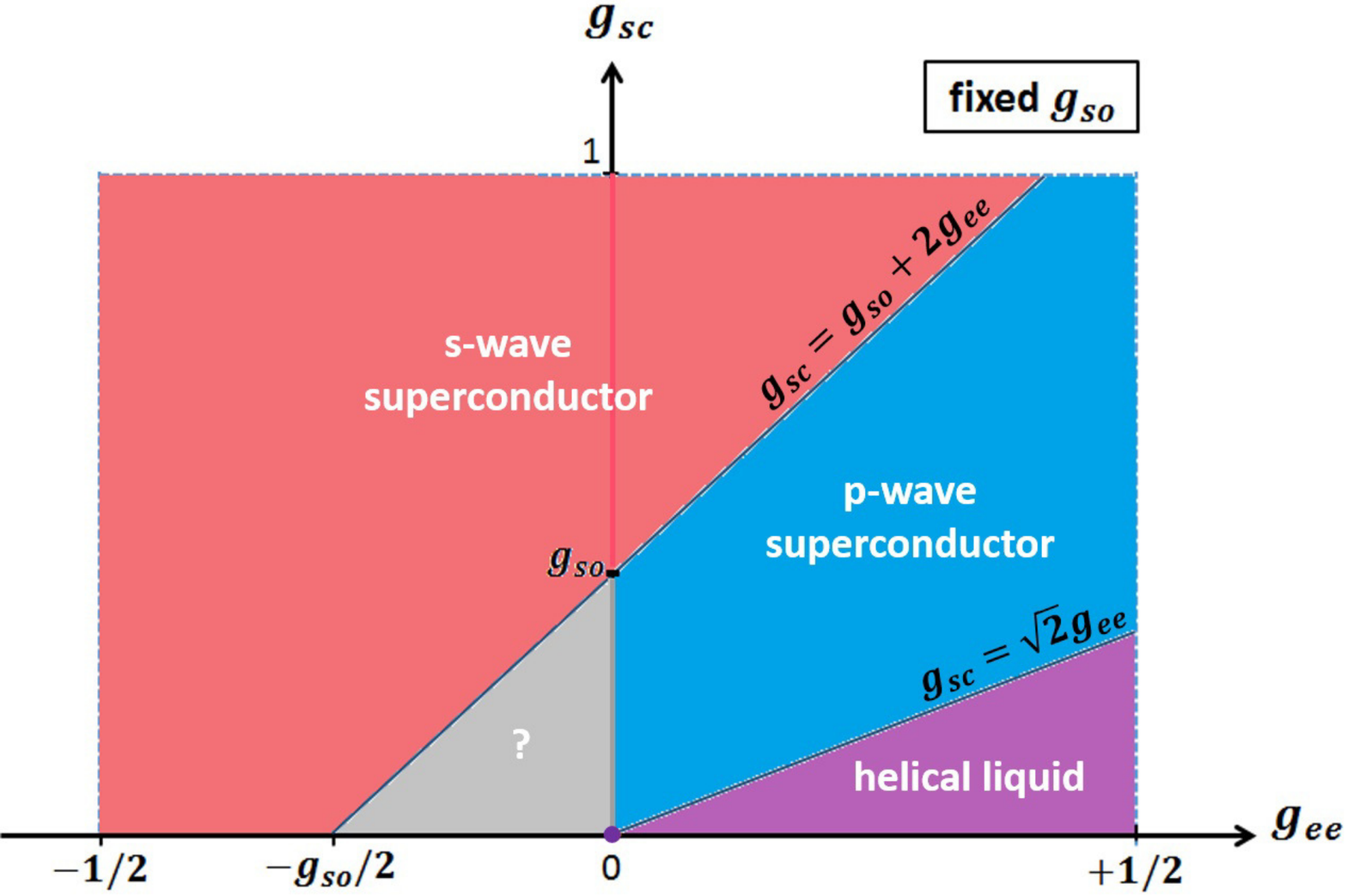}
\caption{(Color online) A fixed-$g_{so}$ cut through the phase diagram of the system. The region labeled by a question mark is out of reach of the present formalism.}
\end{figure}

Now, although necessary, the condition that the parameters belong to the $p$-wave superconducting phase depicted in FIG. 6 is not sufficient to guarantee a ``working" topological superconductor. The corresponding conditions on the parameters must be supplemented by at least three ``practical" criteria: $PC_{1}$ - The insulating gap must exceed the superconducting gap, otherwise it becomes energetically favorable to open a superconducting gap at all four Fermi points, thus loosing the $p$-wave state. This parallels the condition that the Zeeman gap in the more conventional scheme for obtaining a 1D spinless $p$-wave superconducor in a quantum wire must be larger than the proximity gap \cite{AliceaReview}; $PC_{2}$ - The superconducting gap itself must exceed the thermal energy $k_{B}T$ at lab temperatures $T$, so as to withstand thermal leakage; $PC_{3}$ - The physical scaling lengths at which the gaps open up (in the language of RG \cite{Giamarchi}) must not exceed the system's cutoff length. In the case of a defect and impurity-free system (realizable in a cold-atom emulation of a quantum wire, cf. Sec. VI.B), the cutoff length is the system's size $Na$, while for a quantum wire in a semiconductor heterostructure with electron-impurity scattering, it will be the localization length $L_{loc}$.

The superconducting gap $M_{sc}$ and the insulating gap $M_{ins}$ can be computed from the general expression \cite{Giamarchi}
\begin{equation}
M=\Lambda e^{-l^{\star}},
\label{gap}
\end{equation}
where $M$ is the gap, $\Lambda$ is the RG energy cutoff and $l^{\star}$ is the RG scaling length at which the gap opens up, that is, the dimensionless length at which the coupling $-$ $g_{sc}$ or $g_{so}$ $-$ becomes of order unity; call it $l^{\star}_{sc}$ for $M_{sc}$ and $l^{\star}_{ins}$ for $M_{ins}$.

The physical dimensionful scaling length $L$ at which a gap opens is obtained from the corresponding RG scaling length $l^{\star}$ via:
\begin{equation}
L=a\,e^{l^{\star}}.
\label{physlength}
\end{equation}

Using eqs. (\ref{gap}) and (\ref{physlength}), the practical criteria $PC_{1}$, $PC_{2}$ and $PC_{3}$ translate into:
\begin{equation}
PC_{1}\,:\,L_{ins}<L_{pw}
\label{PC1}
\end{equation}
\begin{equation}
PC_{2}\,\&\,PC_{3}\,:\,L_{sc}<\min\{L_{ther},\,L_{cut}\}.
\label{PC2PC3}
\end{equation}
In $PC_{1}$, the length $L_{pw}\equiv L_{sc}/r$, with $r\geq1$, is the upper bound on $L_{ins}$ above which the $p$-wave state is lost. This upper bound follows from demanding that $M_{ins}\geq rM_{sc}$. The parameter $L_{ther}\equiv\Lambda a/(k_{B}T)$ in $PC_{2}\,\&\,PC_{3}$ is a thermal length such that, if $L_{sc}>L_{therm}$, thermal energy starts to destroy the superconducting pairing in the inner branch and, with increasing temperature, also the insulating state in the outer branch. Finally, $L_{cut}\equiv Na\,(L_{loc})$ is the system's cutoff length for a defect and impurity-free system (quantum wire with electron-impurity scattering). FIG. 7 summarizes the conditions on the various length scales as implied by the practical criteria above.
\begin{figure}
\includegraphics[scale=0.32]{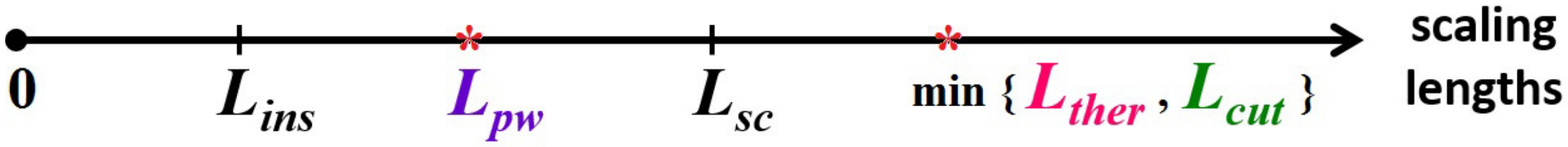}
\caption{(Color online) The scaling lengths $L_{sc}$ and $L_{ins}$ at which the superconducting and insulating gaps, respectively, open up and their corresponding upper bounds $\min\{L_{ther}\,,L_{cut}\}$ and $L_{pw}$ for an observable $p$-wave topological phase.}
\end{figure}

Having established the regime of parameters of an observable $p$-wave superconducting phase, we next present an analysis of the symmetry classes and number of MZMs associated with this topological phase.

\section{Symmetry classes and number of unpaired Majorana zero modes}

As detailed in the previous section, the emergence of a topological superconducting phase in the inner branch - preconditioned by a decoupling of the inner and outer branches in Eqs. (\ref{Houterboson})-(\ref{Hmix}) - requires that the e-e interaction parameter $g_{ee}$ takes on a positive value (or, if a marginally relevant spin-orbit interaction can pull off the decoupling, a value $>-g_{so}/2$), cf. FIG. 5. As a Gedankenexperiment, however, let us temporarily remove the outer branch from the problem entirely, allowing for the emergence of a spinless $p$-wave superconducting phase for any $g_{ee}<0$, that is for $K>1/2$, for which the pairing potential in Eq. (\ref{Hinnerboson}) is strongly RG-relevant. It is then instructive to consider the noninteracting case, $K=1$, for which the effective theory in Eq. (\ref{Hinnerboson}) is simply the bosonized version of the fermionic Hamiltonian ${\cal H}_{inner}$ in Eq. (\ref{Hinner}). This fermionic theory has a linearized spectrum, and, as concerns its topological properties, does not easily fit into the usual topological classification scheme \cite{Schnyder,KitaevClass} since the unboundedness of its spectrum makes the $k$-space topology fuzzy. While a Hamiltonian with a linearized spectrum around the Fermi points may still allow for the identification of differences in the winding numbers which define the 1D topological invariants for different parametrizations \cite{spanslatt2015,carr2015}, it does not {\em per se} provide information about e.g. the number of end-MZMs. For this one needs a Bogoliubov-de Gennes (BdG) Hamiltonian defined on the full Brillouin zone. In other words, the number of unpaired MZMs hosted by ${\cal H}_{inner}+{\cal H}_{\text{e-e}}$ when ${\cal H}_{outer}$ is gapped out can only be deduced from the full underlying theory which exhibits the topological phase. However, in the present case we do not know this theory since the possible topological phase is generated dynamically, and - within our approach - can be accessed only by going to an effective low-energy description with a linearized spectrum. To evade this quagmire we take resort to a construction that connects our low-energy effective theory to an auxiliary theory with a well-defined topological invariant.

For this purpose, let us consider the Hamiltonian which describes, in the full Brillouin zone, a 1D spinless $p$-wave superconductor:
\begin{equation}
H_{p} \!=\! \int \!dx\, \psi\left(-\frac{\partial_x^2}{2m}\! -\! \mu \right)\psi \!- \!\Delta_p \psi \left(\frac{i\partial_x}{k_F}\right) \psi \!+\! \mbox{H.c.}\,,
\label{pwave}
\end{equation}
where $\psi$ is a spinless fermion field, $m$ is an effective mass, $\mu$ is the chemical potential, and $\Delta_p$ is the $p$-wave pairing potential. By linearizing the spectrum of the theory described by Eq. (\ref{pwave}) for $\mu>0$ and writing $\psi(x)=e^{ik_{F}x}R_{+}(x)+e^{-ik_{F}x}L_{-}(x)$, using that to leading order $-i\psi(x)\partial_x\psi(x)\approx2L(x)R(x)$ (and dropping RG irrelevant terms and terms which fluctuate fast and average to zero upon integration), $H_{p}$ in Eq. (\ref{pwave}) gets mapped onto $\int dx\, {\cal H}_{inner}$, with ${\cal H}_{inner}$ given in Eq. (\ref{Hinner}) and $\Delta \approx 2\Delta_p$. An analysis of the BdG Hamiltonian corresponding to $H_{p}$ shows that its band structure has topological winding number $W=1$ for $\mu>0$ and $\Delta$ real-valued \cite{TewariSau}, implying that each end of the wire hosts a single unpaired MZM. Through the mapping above, the same conclusion can be extended to ${\cal H}_{inner}$ with an added e-e interaction provided that the superconducting gap remains finite since, in this case, the unpaired MZMs are known to be stable against e-e interactions \cite{Stoudenmire,SAR,Gangadharaiah,Manolescu,Katsura,Ghazaryan,Gergs}. It follows that if we reverse the procedure and start with a linearized 1D theory supporting a superconducting phase {\em with} e-e interactions, this phase can be smoothly connected to the noninteracting 1D $p$-wave superconductor in Eq. (\ref{pwave}) with well-defined topological properties: Inside the $p$-wave superconducting phase, the e-e interaction is constrained to a ``window of opportunity" so as to stabilize the system of $s$-wave paired helical electrons in an effectively spinless $p$-wave topological phase. This establishes that our scheme is capable of producing a single unpaired MZM at each end of the wire.

The discussion above assumed that the pairing field in Eq. (\ref{pwave}) is constant, with a complex phase which can be gauged away. This puts the topological superconductor in the BDI symmetry class \cite{TewariSau} with manifest time-reversal symmetry, and with a $\mathbb{Z}$ topological invariant calculated as a winding number $W$ which counts the number of unpaired end-MZMs. However, the appearance of a SDW in the outer insulating branch (cf. Sec. III.B) may possibly act as a dynamically generated staggered magnetic field in the topological sector, inducing a phase gradient in the pairing field. Since the inner branch is effectively spinless, this staggered field cannot ``by itself" break time-reversal symmetry in the topological sector. However, it is conceivable that some secondary process could have this effect. (For a case in point, a supercurrent flowing in the bulk superconductor in the proximity to the quantum wire induces a phase gradient in the pairing field \cite{Romito}.) As a result, the symmetry class would then change to D \cite{budich2013}, with a $\mathbb{Z}_2$ topological invariant taking the value unity when $\mu >0$. As before, this implies a single unpaired end-MZMs. While our formalism is not sufficiently powerful to decide whether or not time-reversal symmetry is broken also in the inner branch, the issue is immaterial to our objective to show the emergence of MZMs. In either case, with (or without) symmetry breaking in the inner branch, the D (BDI) symmetry class (with $W=1$) rules that there will be a single unpaired MZM at each end of the wire.

Although not important for our study, one should still keep in mind that the $\mathbb{Z}$ topological invariant of a 1D noninteracting BDI phase gets broken down to $\mathbb{Z}_8$ in presence of interactions, leaving eight distinct equivalence classes \cite{turner2011,fidkowski2011} that can be matched to eight of the ten Altland-Zirnbauer symmetry classes \cite{altland}. While a vital result which highlights the shortcoming of topological band theory for interacting systems, we bypass it by showing that the underlying auxiliary theory without interactions, $H_p$ in Eq. (\ref{pwave}), has a topological winding number with value unity, implying single unpaired end-MZMs for which we can then refer to the stability analyses carried out in Refs. \onlinecite{Gangadharaiah,SAR,Stoudenmire}.

Next, we will attach experimental values to the parameters in order to evaluate the viability of our proposal in light of the theoretical and practical criteria established in Sec. IV. We present two case studies: a quantum wire in a semiconductor quantum well and a quantum wire made of cold atoms trapped in an optical lattice.

\section{Case studies}

\subsection{Case study I: InAs quantum wire}

As a first case study, we investigate the setup of FIG. 1 with the quantum wire patterned in an InAs quantum well (QW).

Starting with the practical criteria encoded in (\ref{PC2PC3}), we may write $L_{therm}=\Lambda a/(k_{B}T)=\hbar v/(k_{B}T)$, with $v$ the drift velocity of the electrons in the semiconductor QW. Using $v\approx10^{5}$ m/s \cite{PB} and $T\approx0.1$ K, which is well above the low temperatures at which the experimental searches for MZMs have been carried out \cite{mourik,deng2012,chang2014}, we get $L_{therm}\approx7.6\,\mu\mbox{m}$. We expect $L_{loc}\approx10\,\mu\mbox{m}$, guided by a prediction by Liu and Das Sarma \cite{liu1995} that the localization length in a high-quality GaAs quantum wire can be several microns long, and using that the electron mobility in an InAs wire is at least 5 times larger than that of a GaAs wire \cite{joyce2013}. Thus, $PC_{2}\&PC_{3}$ demands that $L_{sc}\lesssim7.6\,\mu\mbox{m}$. Using Eq. (\ref{physlength}) with $a \approx 5$ \AA \cite{PB}, the previous condition can be written in terms of the corresponding dimensionless RG scaling length as $l^{\star}_{sc}\lesssim9.6$. 
It remains to check whether this condition is experimentally attainable.

According to the analysis from Section IV, the $p$-wave superconductor is expected to exist in the bare $g_{ee}>0$ side of the phase diagram (see FIG. 6). We take the bare e-e interaction parameter $g_{ee}=0.1$ and, with this choice, we then ask what is the lower bound on the bare superconducting parameter $g_{sc}$ so that its RG flow reaches unity at a scaling length $l^{\star}_{sc}\lesssim9.6$. We find that $g_{sc}\gtrsim0.17$, with a larger $g_{ee}$ implying a larger lower bound on $g_{sc}$.

Turning to the practical criterion (\ref{PC1}), maximizing the upper bound on $L_{ins}$ by taking $L_{sc}\approx7.6\,\mu\mbox{m}$ (corresponding to our choice $g_{ee}=0.1$ with $g_{sc}\approx0.17$) and taking $r=2$ (corresponding to an insulating gap at least twice as large as the superconducting gap), we obtain that $L_{ins}\lesssim3.8\,\mu\mbox{m}$ or, using Eq. (\ref{physlength}), $l^{\star}_{ins}\lesssim8.9$. It is interesting to translate the condition on the length scale at which the insulating gap opens in terms of an effective magnetic field to compare with the value from a magnetic field-assisted topological superconductor. Our choice above of making $L_{sc}\approx L_{therm}$ corresponds to requiring an insulating gap $M_{ins}>2M_{sc}\approx2k_{B}T$. Equating $M_{ins}$ and the Zeeman energy $g\mu_{B}B/2$ of a spin-orbit coupled quantum wire subject to a magnetic field $B$, we get an effective $B>30$ mT at $T\approx0.1$ K, with $g\approx20$ for the $g$-factor in an InAs wire in proximity to an aluminium superconductor\cite{Das}. We note that our lower bound on $B$ is in agreement with the value of 50 mT for which a sharp zero bias peak (associated with the transition to the topological phase and hence the appearance of Majorana modes) is observed in Ref. [\onlinecite{Das}].

We now search for the lower bound on the bare value of the spin-orbit parameter $g_{so}$ so that, under RG, $g_{so}$ approaches unity at a scaling length $l^{\star}_{ins}\lesssim8.9$. We find $g_{so}\gtrsim0.023$, with a larger $g_{sc}$ or a smaller $g_{ee}$ corresponding to a larger lower bound on $g_{so}$.

Since $K\approx1/2-g_{ee}$, $g_{ee}=0.1$ requires a Luttinger parameter $K\approx0.4$. Recalling that $K=(1+g_{2}/(\pi v_{F}))^{-1/2}$, the value of $K$ can be adjusted via the intensity of the $g_{2}$-scattering by modifying the screening from the dielectrics, the surface of the nearby superconductor, and the metallic electrodes attached to the heterostructure which defines the quantum well. Focussing on metallic screening, a detailed analysis \cite{ByczukDietl} shows that, to leading order,
\begin{equation} \label{Cinteraction}
g_{2}\approx\frac{e^2}{\pi \epsilon_0 \epsilon_r} \ln
(\frac{2d}{\xi}),
\end{equation}
where $\xi$ is the width of the quantum wire and $\epsilon_r$ is the averaged relative permittivity of the dopant and capping layers between the quantum well and the nearest metallic surface, at a distance $d$ from the wire. Given this, the setup is now to be designed in such a way that the parameters in Eq. (\ref{Cinteraction}) produce a value of $g_{2}$ corresponding to the desired target value of $K$, with $d$ playing the role of a tuning parameter. We should caution the reader that the bosonization formalism in fact constrains the validity of the expression $K=(1+g_{2}/(\pi v_{F}))^{-1/2}$ used above to the weak-coupling limit $K\approx1$. Still, {\em Bethe Ansatz} and numerical results for this class of models suggest that this expression well captures the effective $K$-parameter also for intermediate to strong coupling \cite{SchulzReview}.

As for the condition on the value of the spin-orbit coupling, $g_{so}\gtrsim0.023$, we now use our previous definitions to write $g_{so}=(2\hbar v)^{-2}\alpha'^{2}/[(\alpha/\beta)^{2}+1]$ where $\alpha=a\gamma_{R}$, $\beta=a\gamma_{D}$, $\alpha'=a\gamma'_{R}$ and $\hbar v=v_{F}$. With $\alpha/\beta\approx2$ (drawn from experimental estimates that $\alpha/\beta$ for a conventionally gated InAs wire is in the range $[1.6,2.3]$ \cite{giglberger2007}) and the same $v$ as above, $g_{so}\gtrsim0.023$ implies that $\alpha'\gtrsim5\times10^{-11}$ eVm. As a point of reference, this may be compared with data from an InAs quantum well capped by a solid PEO/LiClO4 electrolyte, where the Rashba coupling was found to change from $0.4\times10^{-11}$ eVm to $2.8\times10^{-11}$ eVm when tuning a top gate from $0.3$ to $0.8$ V \cite{liang2012}. Thus, our lower bound on $\alpha'$ is around twice as large as the largest experimental value from Ref. \onlinecite{liang2012}. However, the same data reveals a Rashba coupling growing almost five times faster than the gate voltage, within the considered range. Supposing the same rate would be maintained in the next voltage injection, a two-fold boost in $\alpha'$ would be possible by raising the voltage to around 1.3 V. Whereas this gives a first estimate, the actual relation between the Rashba coupling energy and the voltage depends on various microscopic details of the material and setup and, hence, might not be a simple linear one. In any case, the tuning of the Rashba coupling through the amplitude of the modulated electric field is a general feature of the system and can be exploited, possibly in association with other techniques.

Coming to the estimate $g_{sc}\gtrsim0.17$, and recalling that $g_{sc}=a\Delta/(\hbar v)$, gives (with the same values for $a$ and $v$ as above) $\Delta\gtrsim3\times10^{2}$ K. The estimated zero-field proximity gap in an InSb-NbTiN hybridized device \cite{mourik} is $\Delta\approx3.5$ K, with this value being in the upper range of what has so far been reported from experiments. Therefore, our lower bound is two orders of magnitude above the present experimental capability, calling for a material and engineering breakthrough if our scheme is to become viable.


We conclude that a realization of the proposed setup using an InAs quantum well is not feasible with present day technologies and within the regime of parameter values for which our formalism applies, especially in regard to the required strength of the proximity effect. Nonetheless, it is worth noting that should the $p$-wave superconducting phase penetrate into the grey unknown region of FIG. 6, the chances of an experimental realization would be significantly improved. Assuming a proximity pairing in the upper experimental limit ($\Delta\approx3.5$ K leading to $g_{sc}\approx2.3\times10^{-3}$) would require that $g_{ee}\lesssim-0.20$ which, in turn, would demand $g_{so}\gtrsim0.43$. The $g_{ee}$ value can always be adjusted, in principle, through metallic screening, as discussed above. The new lower bound on $g_{so}$ corresponds to $\alpha'\gtrsim19\times10^{-11}$ eVm. Using the same extrapolation as before, the latter value could be attainable by raising the external voltage to around 4 V.

\subsection{Case study II: Spin-orbit-coupled cold atoms}

Observations of $p$-wave Feshbach resonances in spin-polarized $^{40}$K and $^6$Li atoms \cite{regal2003} have spurred hopes that a $p$-wave superfluid of fermionic cold atoms may soon be realized \cite{botelho2005,gurarie2005,cheng2005}. However, the short lifetimes of the $p$-wave pairs in experiments \cite{gaebler2007} make this prospect appear challenging. Various alternative ways of generating a $p$-wave superfluid phase have been proposed, like the one by Zhang {\em et al.} \cite{zhang2008} where an $s$-wave Feshbach resonance is combined with an artificial spin-orbit coupling to produce a 2D $p_x+ip_y$ superfluid. Several other proposals to realize topological phases with cold atoms are discussed in Refs.\,\onlinecite{Sato,jiang2011,goldman2010,stanescu2010,levinsen2011,sun2011,diehl2011,BudicColdAtom}. Could our scheme provide a new vista, now specifically for generating a one-dimensional spinless $p$-wave superfluid exhibiting MZMs? While a precise blueprint for an experimental setup is beyond this work, we shall attempt an analysis of the various components that go into it: (i) a repulsively interacting cold gas of fermionic atoms trapped in a 1D optical lattice; (ii) a uniform coupling to Rashba- and Dresselhaus-type spin-orbit fields; (iii) proximity coupling to a reservoir of $s$-wave paired fermions; and (iv) a spatially modulated Rashba-type spin-orbit interaction.

As for (i), there are by now a multitude of experimental reports of ultracold gases of fermionic atoms confined to one-dimensional lattices \cite{guan2013}. Experiments on $^{40}$K in a 3D optical lattice \cite{jordens2008,schneider2008} have shown that the repulsive interaction strength as measured by $U/t$ (with $U$ a Hubbard-like on-site coupling  and $t$ a hopping amplitude) can be tuned up to two orders of magnitude, using a magnetically controlled Feshbach resonance. We should here point out that while alkali atoms only provide for effective on-site interactions, cold atomic/molecular systems exhibiting long-range interactions are presently under investigation. One promising candidate is ultracold polar molecules that interact via a long-range dipolar potential \cite{micheli,moses}. Very recently, a possibility to realize effective nearest-neighbor interactions from conventional cold-atoms whose bare interaction is on-site has also been considered in Ref. [\onlinecite{wang}].

The second element, (ii), is also expected to be within easy reach, given the experimental progress in manufacturing synthetic gauge fields \cite{goldman2014}. Specifically, Rashba and Dresselhaus spin-orbit couplings of equal strength can be synthesized in the laboratory from two-photon Raman transitions driven by a pair of laser beams \cite{galitski2013}. The technique $-$ with the equal mixture of Rashba and Dresselhaus couplings dictated by symmetry, and known from condensed matter physics as the ``persistent spin-helix symmetry point" \cite{bernevig2006,mross2009} $-$ has been successfully tested with both $^{40}$K and with $^{6}$Li cold atoms \cite{wang2012,cheuk2012}.

Turning to (iii), a proximity-type pairing can be engineered via the coupling of the fermions to a BEC bulk reservoir of Feshbach molecules, as discussed in Ref.\,\onlinecite{jiang2011}. The coupling between the two systems is here transmitted by a pulsed RF field with a Rabi frequency that sets the scale of the effective proximity pairing.

Finally, considering (iv), we note that a theoretical proposal for emulating a {\em position-dependent} Rashba-type interaction for atomic BECs has very recently been put forward by Su {\em et al.} \cite{su2015}. The scheme relies on cyclically laser coupling internal atomic states in an environment where the detuning from resonance depends on the spatial position. In Ref.\,\onlinecite{su2015}, transitions between magnetically split hyperfine states in $^{87}$Rb are detuned from two-photon Raman resonance using a spatially inhomogeneous magnetic field. The same scheme is expected to apply for cyclically coupled states in the two hyperfine manifolds of the fermionic alkali atoms $^{6}$Li and $^{40}$K, $F=1/2, 3/2$ and $F=9/2, 7/2$, respectively, making a realization appear feasible.

With this as a backdrop, let us now check the expediency of a cold-atom emulation by examining the practical conditions in Sec. IV. In the present picture, it is convenient to use Eq. (\ref{physlength}) to rewrite the practical criterion (\ref{PC2PC3}) in terms of the dimensionless RG scaling length as $l^{\star}_{sc}<\min\{\ln(\Lambda/(k_{B}T)),\ln(N)\}$. We then take $T\approx1$ nK as a typical temperature scale in a cold atom setup and assume $\Lambda$ to be of the same order of magnitude as the Fermi energy, which, importing data from Ref.\,\onlinecite{zhang2008}, gives $\Lambda\approx\hbar\times1$ KHz. This yields $\ln(\Lambda/(k_{B}T))\approx2.0$, with $\ln(\Lambda/(k_{B}T))<\ln(N)$ for any number of atoms $N>8$. Our task is now to assess the experimental viability of satisfying $l^{\star}_{sc}\lesssim2.0$ in the proposed setup.

With the same choice as in Sec VI.A, we take $g_{ee}=0.1$, which is expected to be attainable via the Feshbach resonance technique in a system with nearest-neighbor or longer-range interactions \cite{micheli,moses,wang}. We now ask what is the lower bound on the bare $g_{sc}$ so that its RG flow approaches unity at a length $l^{\star}_{sc}\lesssim2.0$. We find that $g_{sc}\gtrsim0.39$, with a larger $g_{ee}$ enhancing the lower bound on $g_{sc}$.

Coming to the practical criterion (\ref{PC1}), this inequality can be rewritten through Eq. (\ref{physlength}) as $l^{\star}_{ins}<l^{\star}_{sc}-\ln(r)$. Taking $l^{\star}_{sc}\approx2.0$ (corresponding to $g_{ee}=0.1$ with $g_{sc}\approx0.39$) and $r=2$, gives $l^{\star}_{ins}\lesssim1.3$. We find that to have $g_{so}$ reaching unity at an RG length $l^{\star}_{ins}\lesssim1.3$, its bare value must satisfy $g_{so}\gtrsim0.38$, with a larger $g_{sc}$ or a smaller $g_{ee}$ corresponding to a larger lower bound on $g_{so}$.

The numerical estimate $g_{sc}\gtrsim0.39$ corresponds to $\Delta\gtrsim\hbar\times0.39$ kHz where we have used that $g_{sc}=\Delta/E_{F}$, with the Fermi energy $E_{F}=v_{F}/a\approx\hbar\times1$ kHz, as before. According to Ref.\,\onlinecite{jiang2011}, an order-of-magnitude estimate for the effective pairing energy yields $\Delta\approx\hbar\times10$ kHz. Hence our lower bound on $\Delta$ is far below the typical experimental value.

From our parametrization we may write $g_{so}=(2E_{F})^{-2}\gamma_{R}'^{2}/[(\gamma_{R}/\gamma_{D})^{2}+1]$. Applying the estimate $g_{so}\gtrsim0.38$, with $\gamma_{R}\approx\gamma_{D}$ and the same $E_{F}$ as before, we estimate that $\gamma_{R}'\gtrsim\hbar\times1.8$ kHz for a working device. From Ref.\,\onlinecite{zhang2008} we learn that the experimental spin-orbit coupling in cold atoms can reach magnitudes up to $\hbar\times10$ kHz. Therefore, our required lower bound on $\gamma_{R}'$ is well within today's capabilities.

It is also interesting to numerically examine the condition $g_{ee}<0$, in case the $p$-wave superfluid pairing does advance into this regime. In fact, in the best studied cold-atom realizations of repulsively interacting fermions using alkali atoms \cite{jordens2008,schneider2008}, the predominantly on-site (Hubbard-like) character of the interaction restricts the Luttinger parameter $K$ to be above 1/2 (hence $g_{ee}$ below 0), with $K=1/2$ corresponding to an infinitely strong on-site repulsion. Assuming a repulsion of intermediary strength we pick $g_{ee}=-0.25$. The resulting conditions for a working device are $g_{sc}\gtrsim0.18$ and $g_{so}\gtrsim0.78$, or $\Delta\gtrsim\hbar\times0.18$ kHz and $\gamma'_{R}\gtrsim\hbar\times2.5$ kHz. Both value are well within the experimental range quoted above.

The numerical estimates above indicate that a sizeable part of the $p$-wave superfluid phase of the cold-atom implementation is expected to be safely within the realm of what is possible to probe in the laboratory.

\section{Summary}

We have proposed and analyzed a magnetic-field-free scheme for synthesizing unpaired Majorana zero modes at the ends of a single-channel quantum wire. In a solid state realization, the wire is modeled as gated by a periodic array of charged top gates, supporting Rashba, Dresselhaus, and e-e interactions, and proximity-coupled to an $s$-wave superconductor which induces a topological $p$-wave superconducting phase. This type of all-electric device for synthesizing Majorana zero modes, if realizable, would be an important step towards applications in topological quantum computing.

The microscopic Hamiltonian which describes the proximity-coupled quantum wire is cast in a low-energy bosonized form that is treated using a renormalization group approach. This formalism allows us to derive the RG flow equations of the theory and predict the phase diagram of the system. Adding ``practical" limits (determined by temperature and by the size of the system) on the RG length scales, we extract the conditions for a working device in the laboratory.

Estimates based on a case study of an InAs wire, contacted to a Nb or Al s-wave superconductor, indicate that a realization of our scheme in a hybrid semiconductor-superconductor device requires improvements upon present-day materials and design capabilities. A variant of our scheme where spin-orbit-coupled ultracold femionic atoms trapped in an optical lattice are effectively proximity-coupled to a BEC reservoir of Feshbach molecules may provide a more easily accessible platform, at least for now. While there already exists a large number of proposals for synthesizing Majorana zero modes using cold fermionic atoms \cite{Sato,jiang2011,goldman2010,stanescu2010,levinsen2011,sun2011,diehl2011,BudicColdAtom}, ours is distinguished by taking advantage of a Feshbach-generated repulsive interaction between the atoms. Its realization would be fascinating, opening up an experimental window on how to drive a topological quantum phase transition (from $s$-wave pairing to spinless $p$-wave pairing) by tuning the strength of an effective fermion interaction. \\

{\bf Acknowledgements} We thank E. Ardonne, K. Le Hur, H. Pu, P. Sacramento, A. Tagliacozzo, H.-Q. Xu, and F. Zhang for valuable input. H.J. acknowledges hospitality at CPTH at \' Ecole Polytechnique where part of this work was carried out. This research was supported by CNPq and CAPES (M.M.), Georgian National Science Foundation and Science and Technology Center in Ukraine through the joint grant No. STCU-5893 (G.J.), and the Swedish Research Council and STINT (H.J.)

\end{document}